\documentclass[twocolumn,secnumarabic,amssymb, nobibnotes, aps, prl, superscriptaddress]{revtex4-2}

\usepackage{graphicx}
\usepackage{dcolumn}
\usepackage{bm}
\usepackage{amsmath}
\usepackage{enumerate}
\usepackage{setspace}
\usepackage{dsfont}
\usepackage{subfigure}
\usepackage{multirow}
\usepackage{indentfirst} 
\usepackage {mathrsfs}
\usepackage{color}

\usepackage{lineno}

\usepackage{makecell}   
\usepackage{multirow}	

\usepackage{threeparttable}  
\usepackage[pdfstartview=FitH,
CJKbookmarks=true,
bookmarksnumbered=true,
bookmarksopen=true,
colorlinks, 
pdfborder=001,
linkcolor=blue,
anchorcolor=blue,
citecolor=blue
]{hyperref}

\begin{document}
	

\title{Time-Frequency Transfer over Optical Fiber}%

\author{Ziyang Chen}%
\affiliation{State Key Laboratory of Photonics and Communications, School of Electronics, and Center for Quantum Information Technology, Peking University, Beijing 100871, China}

\author{Yufei Zhang}%
\affiliation{State Key Laboratory of Photonics and Communications, School of Electronics, and Center for Quantum Information Technology, Peking University, Beijing 100871, China}

\author{Bin Luo}%
\affiliation{State Key Laboratory of Information Photonics and Optical Communications, Beijing University of Posts and Telecommunications, Beijing 100876, China}

\author{Hong Guo}%
\email[E-mail: ]{hongguo@pku.edu.cn}
\affiliation{State Key Laboratory of Photonics and Communications, School of Electronics, and Center for Quantum Information Technology, Peking University, Beijing 100871, China}


\begin{abstract} 
Optical time--frequency transfer establishes the metrological linkage in large-scale clock networks, which facilitates various applications. Fiber-based transfer benefits from the abundant deployment of fiber infrastructures to achieve this advantage. In this Review, we provide an overview of the advances in optical two-way time--frequency transfer, which began with characterizing the time--frequency transfer stability. Then, we discuss the system configuration, key modules, main challenges, and mainstream transfer methods. Finally, the Review concludes with an outlook on further applications toward global-scale high-precision clock networks.
\end{abstract}

\maketitle

\section{INTRODUCTION}\label{sec1}

High-precision time-frequency transfer forms the backbone of crucial advancements in global positioning, navigation and timing (PNT)~\cite{JournalofGeodesy_2015,Lewandowski_2011,JournalofGeodesy_2009}, sensitive probes of environmental fluctuations~\cite{doi:10.1126/science.abo1939,doi:10.1126/science.aat4458,Clivati:18}, and fundamental physics testing~\cite{Nature.591.564.2021,PhysRevLett.118.221102,Nat.Commun.7.12443.2016,Nat.Phys.10.933.2014}, where synchronization at the level of $10^{-18}$ seconds or better is increasingly demanded. However, disseminating these ultra-stable signals over long distances faces significant challenges, including environmental disturbances, signal degradation, and infrastructure limitations. While methods using free-space or satellite techniques offer flexibility for mobile and remote applications, their vulnerability to atmospheric disturbances and line-of-sight constraints compromise their stability.

In contrast, optical fiber networks, widely deployed for telecommunications, present a resilient option by combining inherent immunity to environmental disturbances with the capability for active noise elimination. Leveraging existing fiber infrastructure, fiber-based optical two-way time-frequency transfer (O-TWTFT) achieves unparalleled stability across vast distances, enabling applications such as the redefinition of the second~\cite{Leschiutta_2005,Dimarcq_2024}, continent-spanning clock networks, resilient PNT~\cite{critchley-marrows_ensuring_2024,noauthor_analyzing_2021,J.Geodesy.Geoinf.Sci_2019} and precision geodesy. Here, we focus exclusively on fiber-based approaches due to their mature infrastructure and high stability.

\begin{figure*}
	\centering
	\includegraphics[width=0.5 \linewidth]{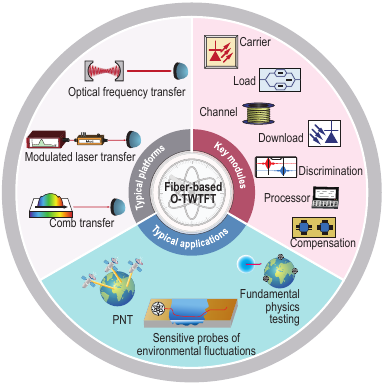}
	\caption{Overview of fiber-based O-TWTFT. The research scope covers the typical platforms, key modules including the carrier, load, channel, download, discrimination, processor, and compensation, and typical applications of O-TWTFT.}\label{Overview_TWTFT}
\end{figure*}

Until now, O-TWTFT over optical fiber links has tremendously progressed. In terms of long-haul scenarios, high-quality transmissions of optical, microwave frequencies and timescale signals have been achieved over thousands of kilometers of cascaded fibers~\cite{schioppo_comparing_2022,droste_optical-frequency_2013,predehl_920-kilometer_2012,yu_microwave_2024,Gao:23,Lin:21,ChinPhysLett.41.064202}. In terms of high-precision scenarios, O-TWTFTs have achieved residual fractional frequency instabilities of the order of $10^{-21}$ in optical fibers~\cite{PhysRevA.90.061802}. Although specific system configurations may vary across different application scenarios, the fundamental architectural concept remains consistent. Fig.~\ref{Overview_TWTFT} shows an overview of typical fiber-based O-TWTFT platforms, key modules, and typical applications.

This Review focuses on the advancement of fiber-based O-TWTFT, which has progressed from laboratory demonstrations to field applications that rival the performance of free-space systems. We begin by providing basic notions of the time–frequency transfer and summarizing the widely used characterization methods for frequency stability from the statistics of time series and phase noise spectrum, followed by an analysis of the general configuration of an O-TWTFT system and key modules. We then evaluate mainstream transfer methods, including optical frequency, modulated laser, and comb-based schemes, highlighting their respective milestones and limitations. Finally, we discuss emerging challenges and opportunities, including the development of novel transfer methods, and integration with free-space networks, toward realizing a globally synchronized metrological infrastructure.

\section{BASIC NOTIONS}\label{sec2}

Optical time--frequency transfer achieves remote sharing of time--frequency signals through optical means. The design of O-TWTFT systems depends on the type of transmitted signals such as optical frequency signals, microwave frequency signals, and timescale signals. In this Review, we define the following basic notions for discussion.

\begin{itemize}
	\item \textbf{Frequency signal.} The frequency describes the number of repeated occurrences of an event per unit of time. The frequency signal in this Review is mathematically denoted as the quasi-perfect sinusoidal signal~\cite{Rubiola_2008}.
	\item \textbf{Timescale signal}. From the user's perspective, time signals should contain two aspects of information: the moment under consideration and the temporal interval between two moments. In most cases, for ease of use, periodic pulse signals are used to denote timescale (or time interval) data, and a specific instance of a pulse is denoted by a timestamp.
	
	\item \textbf{Time--frequency transfer}. This Review focuses on the precision of time--frequency transfer, so the scope is limited to the transmission of frequency or timescale signals using optical means. Timestamps can be distributed using conventional communication methods.
	
	\item 
	\textbf{O-TWTFT}. The basic idea of O-TWTFT is to use optical carriers to carry time--frequency signals and use the reciprocity of bidirectional light travel to cancel most of the common-mode noises and the link drift to reproduce time--frequency signals with high precision at remote locations. 
	
	\item  \textbf{Time--frequency transfer stability}. In the field of precise time and frequency measurement, the quality of a oscillator's output frequency encompasses two main aspects, typically reflected and evaluated by two metrics, namely accuracy and stability. Accuracy reflects the relationship between the actual output frequency and the nominal frequency; stability reflects the variation of the actual output frequency over time. 
	
	Since the accuracy of time-frequency transfer depends on the clock itself, time-frequency transfer is more concerned with the stability added to the clock by the system. The stability of time-frequency transfer refers to the additional stability introduced solely by the time-frequency transfer system during the clock transfer process, excluding the inherent stability of the clock source.

\end{itemize}

\section{CHARACTERIZATION OF TIME-FREQUENCY TRANSFER STABILITY\label{sec3}}

Before we formally begin the discussion, it is necessary to clarify the definitions of various time-frequency metrics and their relationships. a comprehensive and well-founded set of metrics is essential to evaluate the ``quality'' of the transmitted signal, including both accuracy and stability. Comparisons of performance across different approaches are often sought, with a focus on contrasting the diverse experimental metrics that have been reported. However, there are numerous types of time-frequency metrics reported in existing articles, making direct comparisons of different metrics often inappropriate. 

Here, we have outlined the main time-frequency transfer metrics and their relationships, hoping to clarify the ways in which results of different works can be compared. It is noteworthy that some metrics can be approximately compared through certain mathematical relations described in the following text, while others cannot. Therefore, when describing metrics, we posit that raw data and the corresponding \textit{Power Spectral Density} (PSD) can reflect the most fundamental data characteristics, whereas other statistical metrics merely capture a fraction of the statistical information we anticipate from the data. It is essential to choose appropriate statistical metrics to describe the performance of the system based on the specific scenario.

\subsection{Concepts and Models}
The clock signal of oscillators, represented by high-precision quantum frequency standards, can generally be expressed by the quasi-perfect sinusoidal signal~\cite{Rubiola_2008}, given by
\begin{equation}
	v(t)=V_0[1+\alpha(t)]\cos(\Phi(t)),
	\label{oscillator}
\end{equation}
where $V_0$ represents the amplitude of the signal, $\alpha(t)$ represents the relative amplitude jitter, and $\Phi(t)$ represents the phase of the periodic signal, which can be expanded as~\cite{Lindsey1976}
\begin{equation}
	\Phi(t)=\omega_0t+\sum_{k=2}^{k_\text{max}}\frac{\Omega_{k-1}}{k!}t^k+[\psi(t)-\psi(0)]+\Phi(0).
	\label{phase}
\end{equation}
By differentiating with respect to time, we can obtain the instantaneous frequency of the oscillator as
\begin{equation}
	\dot{\Phi}(t)= \omega_0+\sum_{k=1}^{k_\text{max}-1}\frac{\Omega_k}{k!}t^k+\dot{\psi}(t).
	\label{frequency}
\end{equation}
The first term $\omega_0=2\pi f_0$ represents the nominal frequency, namely the center frequency of the oscillator. The second term represents the system disturbance, and $\Omega_k$ is a series of time-independent random variables used to describe the $k^\text{th}$-order frequency drift. It reflects the impact on the system's accuracy, similar to a form of systematic error, which can be mitigated by considering and compensating for a series of physical effects such as Doppler shift, collision shift or by performing higher-order differences on the instantaneous frequency. The third term is the noise term, which is a purely random process and can only be described using statistical methods. It reflects the system's stability, which is also of greater concern in time-frequency transfer. Furthermore, we normalize the phase and instantaneous frequency with respect to the center frequency, yielding~\cite{IEEE2009,Allan1975}
\begin{equation}
	\left\{
	\begin{aligned}
		&x(t)=\frac{\Phi(t)}{2\pi f_0}, [\rm{s}]\\
		&y(t)=\frac{dx(t)}{dt}=\frac{\dot{\Phi}(t)}{2\pi f_0}. [\rm{dimensionless}]
	\end{aligned}
	\right.
	\label{normalized}
\end{equation}
In this case, $x(t)$ is often referred to as phase-time fluctuation in the literature, while $y(t)$ is generally called fractional frequency fluctuation.

In the experiment, we collect the time (phase) or frequency signal to obtain a data run $\{x_i\}$ or $\{y_i\}$, with the minimum sampling interval $\tau_0$. These two signals can also be transformed into each other with the formula
\begin{equation}
	x_{i+1}=x_i+y_i\tau_0.
	\label{xi_yi_conversion}
\end{equation}
Also note that, in the standard case, we denote the total number of data points in $\{x_i\}$ as $N$ and in $\{y_i\}$ as $M$, with $N = M + 1$. 

\subsection{Metrics at a Glance}
After clarifying the basic signal models, concepts, and the transform relationship mentioned above, we further provide a systematic review of the commonly used evaluation metrics in the field of time-frequency transfer. The analysis and evaluation of high-precision time--frequency signals can be divided into two parts in terms of the application scenarios: the frequency domain and the time domain, and the detailed analysis can be seen in Ref.~\cite{Rubiola_2008}. Table~\ref{tab:Metrics_Overview} provides a glance of the metrics that will be introduced in the following text, and the specific calculation formulas are provided in Appendix A.
\begin{table*}[t]
	\vspace*{-2pt}
	\centering
	\caption{A glance of the metrics discussed in this Review. Quantities without explicit units in parentheses are dimensionless metrics. PM: phase modulation.}
	\label{tab:Metrics_Overview}
	\begin{tabular}{ccc}
		\hline\hline
		& \textbf{Metrics} & \textbf{Main functions/Advantages} \\ \hline
		\multirow{2}{*}{\textbf{\begin{tabular}[c]{@{}c@{}}Frequency\\ domain\end{tabular}}} & Power spectral density ($[\text{s}^2/\text{Hz}]$ or $[1/\text{Hz}]$) & \begin{tabular}[c]{@{}c@{}}Represents noise variance density in the frequency domain.\end{tabular} \\& Timing jitter ($[\text{s}]$) & \begin{tabular}[c]{@{}c@{}}Reflects variations of the timing signal from ideal positions.\end{tabular} \\ \hline
		\multirow{9}{*}{\textbf{\begin{tabular}[c]{@{}c@{}}Time\\ domain\end{tabular}}} & \begin{tabular}[c]{@{}c@{}}Allan variance (non-overlapping)\end{tabular} & \begin{tabular}[c]{@{}c@{}}Evaluates second-order-stationary stochastic phase process.\end{tabular} \\
		& Modified Allan variance & \begin{tabular}[c]{@{}c@{}}Distinguishes between white and flicker PM noises.\end{tabular} \\
		& Time variance ($[\text{s}^2]$) & \begin{tabular}[c]{@{}c@{}}Commonly used in evaluating the stability in time transfer.\end{tabular} \\
		& \begin{tabular}[c]{@{}c@{}}Allan variance (overlapping)\end{tabular} & \begin{tabular}[c]{@{}c@{}}Reduces statistical errors by making full use of data.\end{tabular} \\
		& Total variance & \begin{tabular}[c]{@{}c@{}}Further improves the long-term stability evaluation.\end{tabular} \\
		& Theoretical variance \#1 & \begin{tabular}[c]{@{}c@{}}Estimates very long-term stability up to 75\% total data run time.\end{tabular} \\
		& TheoH & \begin{tabular}[c]{@{}c@{}}Optimizes theoretical variance \#1 by eliminating the bias.\end{tabular} \\
		& Hadamard variance & One-order higher difference than the Allan variance. \\
		& High-order structure functions & Handles processes with even higher-order frequency drifts. \\ \hline\hline
	\end{tabular}
\end{table*}

\subsection{Historical Highlights}

\paragraph{\textbf{Frequency domain of the original data}} Historically, the analysis of frequency noise was primarily conducted in the frequency domain.

From the perspective of the frequency domain, the most important metrics are the PSDs of phase time and instantaneous frequency, that is, $S_x(f)~[\text{s}^2/\text{Hz}]$ and $S_y(f)~[1/\text{Hz}]$. Based on Eq.~(\ref{normalized}) and the properties of Fourier transform, they satisfy the relationship
\begin{equation}
	S_y(f) = f^2 S_x(f),
	\label{PSD}
\end{equation}
where $f$ denotes the Fourier frequency, and note that the one-sided PSD is used here and in the following text. 

In general, the PSD on a double logarithmic scale are often ploted, and $\alpha$ and $\beta$ are used to represent the slopes of $S_y(f)$ and $S_x(f)$, respectively. Combining with Eq.~(\ref{PSD}), it is clear that $\alpha = \beta + 2$. Here, different slopes correspond to different types of noise. Specifically, the values $\alpha = 2, 1, 0, -1, -2$ correspond to the five most common types of phase/frequency noise: white phase modulation (PM), flicker PM, white frequency modulation (FM), flicker FM, and random walk FM. This classification of noise types based on PSD is commonly referred to as the power-law model in the literature.

By analyzing the PSD of signal noise, one can obtain the energy of noise at different frequency points, which is used to represent the intensity of noise at different frequencies. Due to the fact that the PSD is obtained through Fourier transform of all original data, it can best reflect the frequency information of the original data (only losing the phase information of the original data in the frequency domain).

Moreover, in scenarios such as laser systems, the focus often lies on timing jitter. Obtained by integrating the PSD within a specific frequency range, the timing jitter reflects variations of the timing signal from ideal positions in time excluding frequency offsets and drifts~\cite{IEEE2009}.

\paragraph{\textbf{First-order stationary process of phase}} Compared to frequency domain analysis, time domain analysis is more intuitive in describing signal fluctuations. For evaluating the stability of a time series, the most straightforward approach is to use the standard variance (SVAR), as it reflects the degree of deviation of a set of data from the mean value. When the phase noise is first-order stationary, the description of SVAR is effective.

\paragraph{\textbf{Second-order stationary process of phase}} In the time-frequency transfer domain, SVAR is nonconvergent for most types of noise and therefore cannot be utilized. 

For example, in a time series containing flicker FM noise, simulations show that as the sample size increases, the SVAR gradually diverges instead of remaining stable. This results in completely different SVAR values calculated from the same stable system when different researchers collect different sample sizes. Clearly, SVAR cannot serve as a reasonable evaluation metric. The root cause of this issue is that SVAR is calculated with respect to the mean value of the data, but in the presence of flicker FM noise, the system exhibits slow-changing fluctuations in the time domain, causing the average of the data run to vary with the sample size.

To address the serious limitations of traditional evaluation metrics, in the 1960s, a series of articles were published by many researchers~\cite{Allan1966, Barnes1966, Lacey1966, Vessot1966}, including D. W. Allan, from National Institute of Standards and Technology (NIST), in journals such as \textit{IEEE Proceedings on Frequency Stability}. Building on the work originally proposed by Allan himself in 1966~\cite{Allan1966}, these discussions culminated in 1971 when the \textit{IEEE Time and Frequency Committee} published a summary that formally defined the two-sample Allan variance (AVAR)~\cite{Barnes1971}, which has been used ever since, namely,
\begin{equation}
	\scalebox{0.95}{$
		\sigma_y^2(\tau)\triangleq\frac{1}{2}\mathbb{E}\left\{\left[\frac{\Phi(t+2\tau)-\Phi(t+\tau)}{\omega_0\tau}-\frac{\Phi(t+\tau)-\Phi(t)}{\omega_0\tau}\right]^2\right\}.
		$}
	\label{2sample}
\end{equation}

From the above definition, we can see that the AVAR no longer involves the expectation of the data sequence, but instead evaluates system stability by calculating the differences between consecutive data points and taking the expectation of their squared values. We can express Eq.~(\ref{2sample}) in a more explicit form at $\tau=\tau_0$ point with normalized phase and instantaneous frequency, i.e.,
\begin{equation}
	\begin{aligned}
		\sigma_y^2(\tau=\tau_0)&=\frac{1}{2(M-1)}\sum_{i=1}^{M-1}(y_{i+1}-y_i)^2\\
		&=\frac{1}{2\tau_0^2(N-2)}\sum_{i=1}^{N-2}(x_{i+2}-2x_{i+1}+x_i)^2.
	\end{aligned}
	\label{allan}
\end{equation}
Thus, it is clear that the essence of the AVAR is the first-order difference of the frequency or the second-order difference of the phase. On the other hand, from the perspective of the structure function~\cite{Lindsey1976}, the AVAR is essentially the first-order structure function of the frequency $D_y^{(1)}$ or the second-order structure function of the phase $D_x^{(2)}$, and Eq.~(\ref{2sample}) can also be written as
\begin{equation}
	\sigma_y^2(\tau)=\frac{D_x^{(2)}}{2\tau^2}=\frac{D_y^{(1)}}{2}.
	\label{AllanDef}
\end{equation}
The detailed definition formula are shown in Appendix A. Therefore, if the phase data follows a second-order stationary stochastic process, AVAR can effectively describe its stability, which is also a key metric of interest for most stable time-frequency systems.

\begin{figure*}[t]
	\centering
	\includegraphics[width=1.0\linewidth]{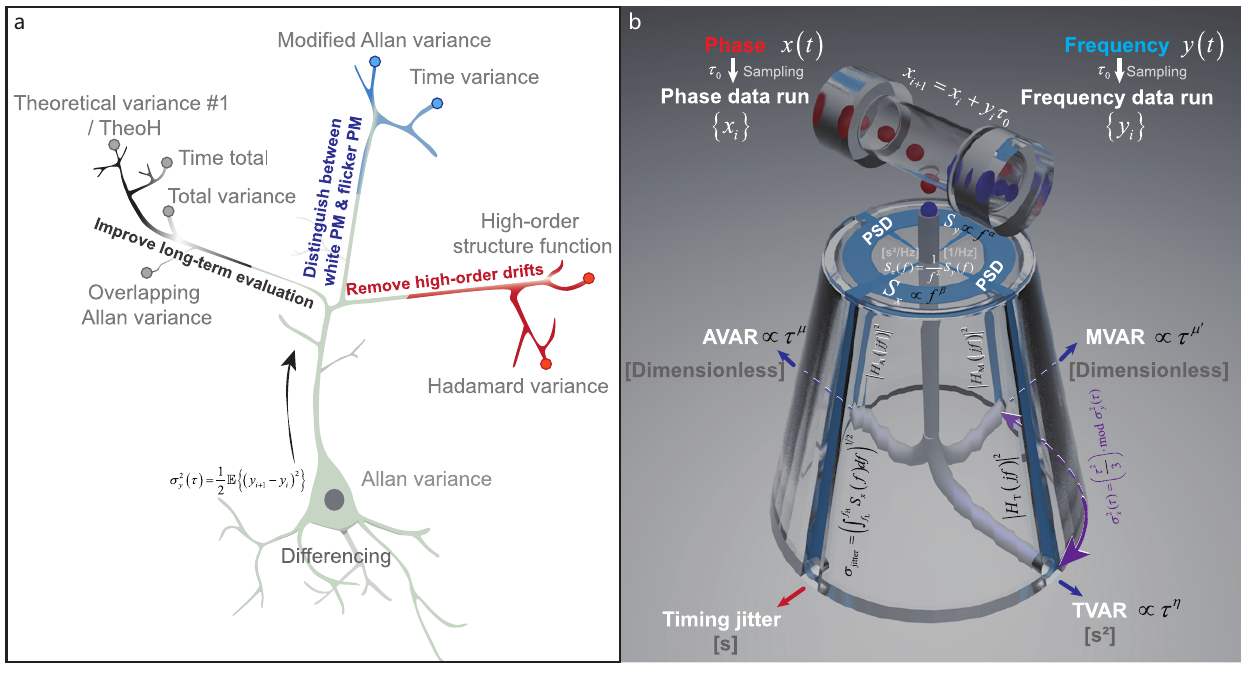}
	\caption{Overview of metrics in time-frequency transfer. a. Time-domain evaluation metrics further developed based on the original AVAR proposed by D. W. Allan~\cite{Allan1966}. Each of these improves upon a specific limitation of the AVAR. b. Five commonly used evaluation metrics in the time-frequency transfer field and their conversion relationships.}
	\label{AllanMainFigure}
\end{figure*}

\paragraph{\textbf{Further development of Allan variance}} The AVAR has evolved into three distinct research branches to address the limitations of AVAR and cater to specific application scenarios. These include Hadamard variance (HVAR), total variance (TOTVAR), theoretical variance \#1 ($\widehat{\text{Theo1}}$) and TheoH, a hybrid statistic as the combination of $\widehat{\text{Theo1}}$ and AVAR. These metrics each represent a modification of the standard Allan variance in some particular aspect. As shown in Fig.~\ref{AllanMainFigure}~a, these improvements can be divided into three branches, each corresponding to a different thought.

(I) \textbf{Distinguish more types of noise.} The first branch is to address the issue that the standard AVAR cannot distinguish certain types of noise. In 1981, D. W. Allan, together with J. A. Barnes, published a paper introducing the modified Allan variance (MVAR)~\cite{Allan1981}. The motivation behind Allan's introduction of this new metric lies in the fact that, as mentioned earlier, different types of noise follow different power laws, and this is also true for the AVAR. In other words, $\sigma_y^2(\tau) \propto \tau^\mu$, where the exponent generally satisfies $\mu = -\alpha - 1$, corresponding to $S_y(f)$. However, for cases where $\alpha \geq 1$, the exponent is always $\mu = -2$, meaning that the AVAR cannot distinguish between the noise types $\alpha = 1$ and $\alpha = 2$. Nevertheless, white PM noise (also known as purple frequency noise) corresponding to $\alpha = 2$ and flicker PM noise (also known as blue frequency noise) corresponding to $\alpha = 1$ are often found to occur simultaneously in quartz crystal oscillators. It is therefore necessary to differentiate between these two, which is the reason for the introduction of the MVAR.

The power law of the MVAR is $\text{mod}~\sigma_y^2(\tau)\propto\tau^{\mu^\prime}$, where $\mu^\prime=-1-\alpha$. In addition to the aforementioned AVAR and MVAR, another commonly used metrics in the field of time-frequency transfer is time variance (TVAR), which has time dimension and is defined based on the MVAR. Also, the power law of the TVAR is $\sigma_x^2(\tau)\propto\tau^{\eta}$, and $\eta=1-\alpha$. The TVAR, especially its square root TDEV, acts as a common-used measure of the stability in time transfer. 

\textbf{(II) Improve long-term evaluation.} The second branch includes overlapping AVAR, TOTVAR, $\widehat{\text{Theo1}}$ and TheoH. These methods address the issue that when using standard AVAR to calculate long-term stability from a set of data acquired over a certain period, the resulting error is large and the confidence needs to be further increased. 

Building on the AVAR, the overlapping AVAR has been further developed. It allows for more efficient reuse of data, thereby reducing statistical errors at the same timescale and improving the confidence of the estimate. 
It is important to note that, due to the significant advantages of the overlapping AVAR over the standard AVAR, the standard AVAR is rarely used. The overlapping AVAR is one of the most commonly used evaluation metrics~\cite{handbook2008,IEEE2009}, and in the literature, it is generally directly denoted as AVAR, without explicitly specifying "overlapping", while Allan deviation (ADEV) represents its arithmetic square root.

However, since standard AVAR has a low data reuse rate, it typically only provides stability corresponding to a timescale $\tau$ within one-tenth of the acquisition time~\cite{IEEE2009}. As a result, obtaining stability over longer timescales becomes quite challenging. To address this issue, D. A. Howe from NIST proposed TOTVAR in 1995~\cite{Howe1995}. This metric requires that the original data be symmetrically flipped at both ends and then concatenated to form a new sequence. Therefore, TOTVAR can calculate long-term stability points at $\tau = (N-1)\tau_0$, and there are always $N-2$ difference results involved in the calculation, which leads to smaller error bars.

In 2003, D. A. Howe proposed $\widehat{\text{Theo1}}$, which made further improvements to the estimation for very long-term stability, specifically designed to address the challenges in this particular application scenario~\cite{Howe2003}. Although TOTVAR had already made significant improvements to the long-term confidence for $\tau$ in the 10\%-50\% range of a data run with total duration $T$, $\widehat{\text{Theo1}}$ aims to be a reliable estimator for ultra-long-term stability intervals greater than $T/2$. Note that the timescale is given by $\tau = 0.75 m \tau_0$, indicating that the very-long-term stability can be extended up to three-quarters of the total data run time.

Moreover, the bias of $\widehat{\text{Theo1}}$ with respect to the AVAR can be defined with the corresponding parameters provided in the reference~\cite{Howe2004} for different noise types. It describes the slope of $\widehat{\text{Theo1}}$ under various noise conditions and shows that $\widehat{\text{Theo1}}$ is unbiased only for white FM noise. To further eliminate this bias, D. A. Howe proposed TheoH in 2004~\cite{Howe2004}. Moreover, in the following period, Howe and his collaborators further studied the confidence intervals of $\widehat{\text{Theo1}}$~\cite{Tasset2004}.

\textbf{(III) Remove high-order drifts.} The third branch is the HVAR, which was first proposed by R. A. Baugh in 1971, and the name "Hadamard" comes from the Hadamard transform~\cite{Baugh1971,Pratt1969}. It can be seen that the HVAR essentially corresponds to the third-order difference of phase and the second-order difference of frequency. Its significance lies in that by combining Eq.~(\ref{frequency}), we can see the limitations of using the AVAR, which is based on lower-order differences. For example, if we consider a random frequency process containing second-order drift $\dot{\Phi}(t) = \omega_0 + \Omega_1 t + \frac{\Omega_2 t^2}{2} + \dot{\psi}(t)$, then the first-order difference
$\Delta\dot{\Phi}(t) = \Omega_1 \tau + \frac{\Omega_2}{2} (2\tau t+ \tau^2) + \Delta\dot{\psi}(t)$, which is hidden behind the AVAR, is time-dependent and represents a non-stationary quantity, rendering it unsuitable as a statistical measure of system stability. Hence, it becomes necessary to introduce higher-order difference statistics, and the HVAR is one such representative. The HVAR is frequently used in the stability analysis of rubidium atomic clocks to eliminate the effects of linear frequency drift and is also particularly useful for analyzing GPS clocks~\cite{handbook2008}. 

Besides the HVAR, evaluation metrics based on higher-order structure functions, namely Eq.~(\ref{StructureFunction}), can also be defined. The definition is similar to Eq.~(\ref{AllanDef}), and the new metric can deal with higher-order frequency drift and handle divergent noise.

In addition to the series of metrics derived from three different approaches to improving AVAR, which were mentioned earlier, some more specialized metrics were generated through combinations of these metrics. These include the modified total variance~\cite{Howe1999}, which combines the features of the MVAR and TOTVAR, allows the MVAR to achieve better confidence in long-term stability; the overlapping Hadamard variance~\cite{handbook2008} and the modified Hadamard variance~\cite{Bregni2006}, which further improve the HVAR; and the Hadamard total variance~\cite{Howe2006,DavidHowe2001}, integrating multiple advantages of the HVAR and TOTVAR.

\subsection{The Relationship between Common Evaluation Metrics}
Figure~\ref{AllanMainFigure}~b illustrates the process of calculating five commonly used system stability evaluation metrics in the time-frequency transfer domain, derived from the original phase or frequency data, as discussed earlier. The red and blue balls represent the phase and frequency data runs collected at the sampling time $\tau_0$, respectively. These can be converted to each other using Eq.~(\ref{xi_yi_conversion}), after which they enter the “metrics bottle" for computation. 

First, starting from the original data, the metrics AVAR, MVAR, and TVAR can be directly computed using Eqs.~(\ref{MVAR}, \ref{TVAR}, \ref{OAVAR}) in Appendix A, as shown by the “pipes" inside the “metrics bottle". Second, the PSD derived from the original data involves a Fourier transform, which is straightforward, and the timing jitter can be directly calculated from the PSD based on its definition. Furthermore, by combining Eq.~(\ref{PSDtoVAR}) in Appendix B, AVAR, MVAR, and TVAR can be derived through integration. However, it is important to note that this step is also an approximation, because the transfer function in Eq.~(\ref{PSDtoVAR}) is the theoretical value when $\tau = m \tau_0 \to \infty$. In practice, this approximation is generally sufficient for applications when $m \geq 10$. Third, the TVAR and MVAR only differ by the factor $\tau^2/3$ according to their definitions, so they can be converted into each other. Depending on the application scenario, researchers in different studies often use one or several of the metrics mentioned above.

However, although the theoretical possibility of converting the time-domain evaluation metric, variance, to the frequency-domain PSD has been discussed in the literature with Mellin transforms~\cite{Lindsey1976}, this transform is generally not achievable~\cite{Greenhall1998}. Therefore, the metrics at the “bottom of the bottle" in Fig.~\ref{AllanMainFigure}~b cannot "move upward" and be converted into the PSD. Furthermore, aside from the purple bidirectional arrows, the metrics at the "bottom of the bottle" cannot be converted into each other. Nevertheless, in practice, we can analyze the noise type through slope analysis and use Table~\ref{NoiseTable} in Appendix B to roughly estimate the conversion between the AVAR and MVAR, thereby partially addressing the issue of inconsistent metrics in the literature, which makes direct comparisons difficult. Moreover, from Table~\ref{NoiseTable}, it can be seen that, in most cases, the value of the MVAR is slightly lower than that of the AVAR.

In summary, among the above metrics, although the most commonly used in the literature is $\sigma_y^2 (\tau)$ or $\text{mod}~\sigma_y^2(\tau)$, in fact, the PSD contains the greatest "amount of information" from the original data and should be placed in the central position. It is strongly recommended to include it in the literature. However, in the absence of the original data, it is generally hard to achieve error-free conversion between different metrics.

\section{SYSTEM CONFIGURATION AND KEY MODULES}\label{sec4}

\begin{figure*}[t]
	\centering
	\includegraphics[width= \linewidth]{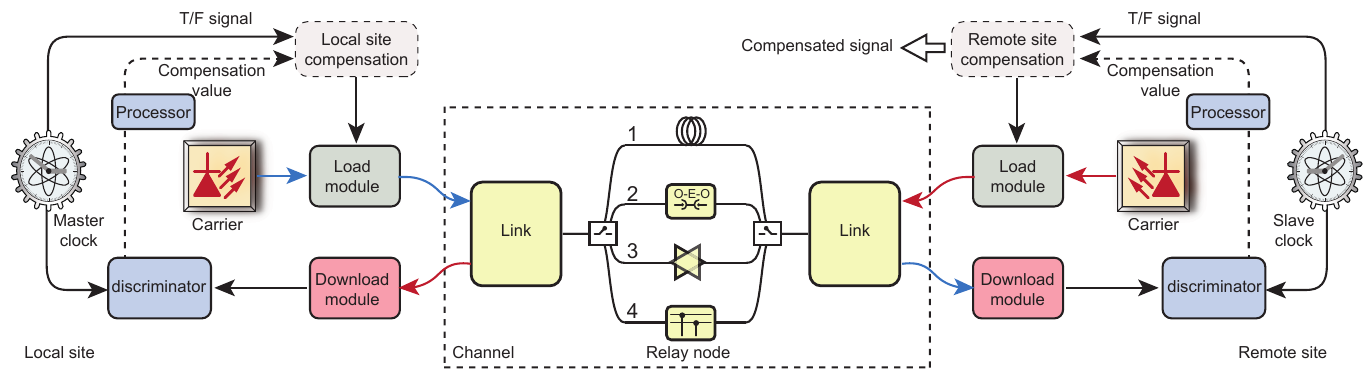}
	\caption{General configuration of an O-TWTFT system. In addition to fiber links, multiplexing and relaying are key technologies in the channel. We have summarized four main relay structures: 1. full loss fiber (also known as the non-relay scenarios); 2. cascaded optical-electro-optical (O-E-O) relay; 3. cascaded all-optical relay; 4. downloadable node. }\label{System_configuration}
\end{figure*}

This section provides an overview of the general configuration and details of the key modules in an O-TWTFT system. Although many O-TWTFT methods have been reported, they can generally be summarized by the architecture in Fig.~\ref{System_configuration}.

The goal of O-TWTFT is to replicate the precision of the master clock at the local site onto the slave clock at the remote site. According to Fig.~\ref{System_configuration}, O-TWTFT can be summarized into the following key components:

\begin{itemize}
	\item \textbf{Master clock.} A locally employed high-precision clock source to produce time--frequency signals. The output can consist of optical frequency signals ($\sim {10^{14}}$ Hz), microwave frequency signals (spanning the general microwave frequency range from kHz to GHz), or timescale signals.
	
	\item \textbf{Slave clock.} A distant clock to be synchronized. In some scenarios such as a round-trip frequency transfer setup with no clock at the remote location, this clock can be provided by the local site transmission.
	
	\item \textbf{Carrier.} Optical carriers to convey time--frequency signals, including lasers (such as narrow linewidth continuous-wave (CW) lasers and modulated lasers) and optical frequency combs, which are often abbreviated as combs.
	
	\item \textbf{Load module.} Methods such as phase locking (stabilization) and modulation to load high-precision time--frequency signals onto optical carriers.
	
	\item \textbf{Channel.} The media to transmit optical signals, which consist of free-space links and optical-fiber links. This Review concentrates on the scenario of optical fibers. In addition to fiber links, multiplexing and relaying are key technologies in the channel. We have summarized four main relay structures: 1. full-loss fiber (also known as the non-relay scenario); 2. cascaded optical-electro-optical (O-E-O) relay; 3. cascaded all-optical relay; 4. downloadable node.   
	\item \textbf{Download module.} The module to recover time--frequency signals with high precision on the received carriers.
	
	\item \textbf{Discriminator.} The component to determine the time or phase difference between the local reference clock and the received signal and facilitate the evaluation of the accumulated noise in the link.
	
	\item \textbf{Processor.} The component to compute and analyze the discriminated data to derive the phase or time values for compensation and subsequently supply them to the compensation module. Typically, in a round-trip scheme, the phase value that must be compensated is denoted as ${{{\varphi _{{\rm{RT}}}}} \mathord{\left/	{\vphantom {{{\varphi _{{\rm{RT}}}}} 2}} \right.			\kern-\nulldelimiterspace} 2}$; in a two-way time transfer, the time difference requiring compensation is denoted as ${{\left( {{t_{\rm{A}}} - {t_{\rm{B}}}} \right)} \mathord{\left/	{\vphantom {{\left( {{t_{\rm{A}}} - {t_{\rm{B}}}} \right)} 2}} \right.\kern-\nulldelimiterspace} 2}$, where ${{\varphi _{{\rm{RT}}}}}$ is the round-trip phase fluctuation, and ${{t_{\rm{A}}}}$ and ${{t_{\rm{B}}}}$ denote the time differences at both sites.
	
	\item \textbf{Local site compensation.} Compensation schemes that implement pre-compensation for fiber-accumulated noise using compensation devices at the local site, such as the feedback and conjugator schemes.
	
	\item \textbf{Remote site compensation.} Compensation schemes that directly compensate the phase noise of the output signals at the remote site, such as the feed-forward scheme. 
\end{itemize}

\section{MAIN CHALLENGES}
Fiber-based O-TWTFT must primarily address a series of issues, including recovery of carrier impairment caused by fiber transmission, cancellation of accumulated noise, especially link-induced phase noise, and implementation hurdles.

\paragraph{\textbf{Signal quality recovery}} The first task is essential to better extract the time--frequency information from a carrier signal. This task involves restoring the energy and shape of the carrier signal.

The energy can be recovered through the structures of relay nodes 2 and 3 in Fig.~\ref{System_configuration}, including the O-E-O relay and all-optical relay. In the O-E-O relay, the noisy optical signal is converted into an electrical signal, which is subsequently reshaped and reconverted into an optical signal whose signal-to-noise ratio (SNR) is improved by the laser transmitter~\cite{Gao:23,9166557}. In addition, a long single span is split into a cascaded multi-span structure to compensate for shorter transmission delays in each split span to overcome the accumulated phase noise over long distances~\cite{rs13112182,gao_dissemination_2015,5361538}.

In the all-optical relay, optical amplifiers such as the erbium-doped fiber amplifier (EDFA)~\cite{6290391} and fiber Brillouin amplifier (FBA)~\cite{droste_optical-frequency_2013} are the key components. In all-optical amplification transmission, since no optical--electrical conversion is involved, new types of noise associated with such conversions are not introduced to the optical signal. Moreover, all-optical nodes eliminate the need to reintroduce new transceivers, which significantly reduces cost, especially when there are numerous cascaded nodes. The advantages and disadvantages of all-optical relay and O-E-O relay are summarized in Table~\ref{tab:Comparison-relay}.

\begin{table*}[t]
	\vspace*{-2pt}
	\centering
	\caption{Comparison of  difference relay methods.}
	\label{tab:Comparison-relay}
	\begin{tabular}{ccc}
		\hline
		\hline
		Relay structure                                                               & All-optical relay                                                                        & O-E-O relay                                                                                   \\ \hline
		\multirow{2}{*}{Dominant noise}                                               & \multirow{2}{*}{Accumulated ASE over spans}                                              & No need for additional transceivers                                                           \\ \cline{3-3} 
		&                                                                                          & Electronic $1/f$ flicker noise                                                                  \\ \hline
		Compensation bandwidth                                                        & \begin{tabular}[c]{@{}c@{}}Smaller (Due to long \\ compensation delay)\end{tabular}     & \begin{tabular}[c]{@{}c@{}}Larger (Due to short \\ segment compensation span)\end{tabular} \\ \hline
		Delay-unsuppressed noise                                                      & Larger                                                                                    & Smaller                                                                                        \\ \hline
		Induced non-reciprocity                                                       & \begin{tabular}[c]{@{}c@{}}Introduced by \\ bidirectional optical amplifier\end{tabular} & \begin{tabular}[c]{@{}c@{}}Introduced by  transceivers\end{tabular}                         \\ \hline
		Signal reshaping                                                              & Difficult to recover large signal distortion                                             & Restore signal quality through resending                                                      \\ \hline
		\begin{tabular}[c]{@{}c@{}}Cost and \\ infrastructure complexity\end{tabular} & No need for additional transceivers                                                      & More transceiver components                                                                   \\ \hline
		Reliability                                                                   & High                                                                                     & More fragile due to the cascaded PLL                                                          \\ 
		\hline
		\hline
	\end{tabular}
\end{table*}  \vspace*{-3pt}

Recovering the signal shape is particularly important in pulse transmission scenarios, such as pulse-modulated time transfer and comb transfer. Pulse shaping and filtering are the commonly used solutions.

We note that, in comb-based time-frequency transfer, signal quality recovery is critical for maintaining high temporal resolution of optical frequency combs over long fiber links. Unlike simple modulated CW laser transfer, combs require additional considerations due to their broadband nature. 

\begin{itemize}
	\item \textbf{Optical filtering.} To minimize the dispersion of the comb while maximizing the utilization of commercial optical communication components for transmission, the bandwidth of the comb is confined within the standard channels of wavelength division multiplexing (WDM) through optical filtering as much as possible.
	
	\item \textbf{Optical amplification.} EDFAs or Raman amplification compensate for fiber attenuation but must be carefully designed to avoid nonlinearities that distort the comb spectrum. Bidirectional EDFA minimizes asymmetry-induced noise in two-way transfer schemes. Low-noise amplifiers (such as FBAs) serve as excellent candidates for selectively amplifing comb lines while suppressing broadband noise~\cite{droste_optical-frequency_2013}.

	\item \textbf{Dispersion compensation.}
	Dispersion compensation fiber (DCF) is used to compensate for the residual dispersion in a fine-tuned manner~\cite{chen_dual-comb-enhanced_2024}. For more precise dispersion compensation, techniques such as liquid crystal modulation could be used~\cite{Opt.Lett.23.283.1998}. Optionally, the impact of dispersion can also be mitigated through hollow-core fibers~\cite{PMID:26490424,feng_stable_2022}.

	\item \textbf{Pulse regeneration and reshaping.} Compensating for severe waveform distortions of combs presents significant challenges. The most effective solution involves implementing high-precision phase locking of the comb's phase information to a new comb, thereby achieving pulse regeneration and reshaping~\cite{yu_microwave_2024}.
\end{itemize}

\paragraph{\textbf{Noise cancellation}} The total residual noise of an O-TWTFT is 
\begin{equation}
	{\varphi _{{\rm{residual}}}} = {\varphi _{{\rm{load}}}} + {\varphi _{{\rm{link}}}} + {\varphi _{{\rm{meas}}{\rm{.}}}},
\end{equation}
where ${\varphi _{{\rm{load}}}}$ and ${\varphi _{{\rm{meas}}{\rm{.}}}}$ are the noises from signal loading and measurement processes, respectively; ${\varphi _{{\rm{link}}}}$ is the residual link-induced noise, which cannot be canceled by two-way propagation or noise compensation techniques.

(I) \textbf{Loading noise.} Since the noise introduced by loading and measurement processes cannot be eliminated through link noise compensation, it must be minimized. Technologies such as fiber-loop optical-microwave phase detection (FLOM-PD)~\cite{Kim:04,Jung:12} and laser frequency stabilization can greatly enhance the capability of carrier tracking references. However, measurement noise optimization remains complex.

(II) \textbf{Measurement noise.} In frequency transfer, the noise of the phase discriminator (such as a mixer) for phase detection is often much lower than the noise in the link itself. Therefore, the primary focus in frequency transfer is to suppress the accumulation of noise from the link. In time transfer, since the time information is extracted from the pulse delay (using devices such as a time interval counter), the time measurement capability is often less precise than directly comparing phases. Thus, the challenge in high-precision time transfer lies in optimizing the pulse measurement capability.

(III) \textbf{Link noise.} Many factors can affect the phase noise of optical fiber links, including chromatic dispersion~\cite{lopez_86-km_2008}, polarization-mode dispersion~\cite{6702214}, temperature, and total delay. Ref.~\cite{MeasurementScienceTechnology_2010} provides a detailed discussion. We note that the two most critical impacts on long-distance fiber transfer are the temperature and delay-unsuppressed noise~\cite{williams_high-stability_2008,newbury_coherent_2007}, which mainly affect the long-term stability and short-term stability of the system, respectively.

We assume that the fiber delay is given by $\tau  = {{nL} \mathord{\left/	{\vphantom {{nL} c}} \right.\kern-\nulldelimiterspace} c}$, where $n$ is the group refractive index and $L$ is the total fiber length. Then, the delay variation is:
\begin{align}
	\Delta\tau  = \left( {\frac{1}{c}\frac{{\partial n}}{{\partial T}} + \frac{n}{c}{\alpha _{{\rm{th}}}}} \right)\Delta T\left( t \right)L + D\Delta \lambda \left( t \right)L,
	\label{delay_fluctuation}
\end{align}
where $\Delta T\left( t \right)$ is the temperature change, ${\alpha _{{\rm{th}}}} = {{\left( {{{\partial L} \mathord{\left/{\vphantom {{\partial L} {\partial T}}} \right.\kern-\nulldelimiterspace} {\partial T}}} \right)} \mathord{\left/{\vphantom {{\left( {{{\partial L} \mathord{\left/	{\vphantom {{\partial L} {\partial T}}} \right.\kern-\nulldelimiterspace} {\partial T}}} \right)} L}} \right.\kern-\nulldelimiterspace} L}$ is the fiber thermal expansion coefficient, $D=\frac{1}{c}(\partial n/\partial\lambda)$ is the chromatic dispersion coefficient, and $\Delta \lambda \left( t \right)$ is the transmitter wavelength, which varies with time. Since the temperature-introduced refractive-index vibration dominates the delay fluctuation~\cite{MeasurementScienceTechnology_2010}, the wavelength-changing-introduced dispersion effect can be ignored. Therefore, we simplify the delay fluctuation by
\begin{equation}\label{Temp_delay}
	\Delta\tau \left( t \right) = \alpha L\Delta T\left( t \right),
\end{equation}
where $\alpha$ is the overall temperature coefficient given by $\alpha  \approx 36.80~{\rm{ps}} \cdot {\rm{k}}{{\rm{m}}^{{\rm{ - 1}}}} \cdot {{\rm{K}}^{{\rm{ - 1}}}}$~\cite{MeasurementScienceTechnology_2010} near the wavelength of 1550 nm. In long-distance transfer, the long-term stability of most systems is dominated by the delay fluctuation noise caused by temperature.

In O-TWTFT, the temperature noise of the long fiber link is a very-low-frequency noise and can easily be suppressed by the feasture of two-way noise cancellation. The biggest factor affecting long-term stability is the temperature noise of asymmetric optical fibers, such as double-path bidirectional optical amplificaitons, local reference path, outloop detection path, et al. Hence, when we use Eq.~(\ref{Temp_delay}) to estimate temperature-induced link fluctuations, $L$ often refers to the length of the asymmetric part of the link.

We note that temperature fluctuations and polarization-mode dispersion remain critical challenges in fiber-based time-frequency transfer, particularly in field trials. Recent advancements in materials science, environmental control systems, and signal processing techniques have enabled promising solutions to these challenges. For instance, adopting emerging fiber architectures like hollow-core fibers has shown significant potential in minimizing temperature-dependent phase noise and polarization variations~\cite{PMID:26490424,feng_stable_2022}. Furthermore, enhanced environmental isolation measures, including deep burial of fiber lines and conduit shielding, are important to reduce daily temperature-induced phase drift. Moreover, state-of-the-art machine learning algorithms offer a novel approach for pre-compensating phase shifts through analysis of historical temperature and phase data to predict drift patterns.

Another key limitation for long-haul transfer is the delay-unsuppressed noise~\cite{williams_high-stability_2008,newbury_coherent_2007}. 
Time-frequency transfer requires compensation for accumulated noise across the entire link, which is achieved through bidirectional or round-trip signal transmission. However, longer links result in fast-varying noise that cannot be fully compensated, which will inevitably introduce the delay-unsuppressed noise. We can simplify the compensation model for time-frequency transfer as a delay-locked loop framework~\cite{8447468}. Consequently, longer links lead to narrower phase-locking bandwidth, resulting in significant residual delay-induced noise.

The tens-of-millisecond link delay disables the full suppression of fiber link noise, and delay-unsuppressed noise remains in the compensated result. According to \cite{williams_high-stability_2008}, the delay-unsuppressed noise can be denoted as
\begin{align}
	{S_{{\rm{unsuppressed}}}}\left( f \right) = \frac{1}{3}{\left( {2\pi f\tau } \right)^2}{S_{{\rm{fiber}}}}\left( f \right) ,
	\label{Unsuppressed_Noise_Eq}
\end{align}
where ${S_{{\rm{fiber}}}}\left( f \right)$ is the free-running link phase noise. The phase noise in
the low-frequency region (below ${1 \mathord{\left/
		{\vphantom {1 {4\tau }}} \right.
		\kern-\nulldelimiterspace} {4\tau }}$) cannot be fully canceled by the compensation system and is limited
by the delay-unsuppressed noise. Furthermore, at higher frequencies, the phase noise is completely uncompensated because the
compensation bandwidth is limited by the link delay $\tau$, given by ${\rm{B}}{{\rm{W}}_{{\rm{compensation}}}} = {1 \mathord{\left/
		{\vphantom {1 {4\tau }}} \right. \kern-\nulldelimiterspace} {4\tau }}$.

\begin{figure}[t]
	\centering
	\includegraphics[width= \linewidth]{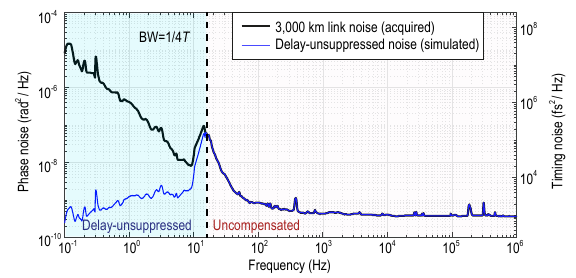}
	\caption{Phase noise spectral density of free running 3000-km link [black curve, acquired by phase analyzer (Microchip, 53100A)] and the simulated delay-unsuppressed noise [blue curve]. The data is from Ref.~\cite{yu_microwave_2024}.
	}\label{delay-unsuppressed}
\end{figure}

\begin{table*}[t]
	\vspace*{-2pt}
	\centering
	\caption{Comparison of lab and field experimental scenarios.}
	\label{tab:Comparison_lab_field }
	\begin{tabular}{ccc}
		\hline\hline
		\begin{tabular}[c]{@{}c@{}}Experimental\\ scenarios\end{tabular} & Lab spooled fiber                                                                                                                    & Field installed fiber                                                                                                                                  \\ \hline
		\multirow{3}{*}{Conditions}                                      & Controlled environment                                                                                                               & \multirow{3}{*}{\begin{tabular}[c]{@{}c@{}}Real-world, exposed (buried/aerial) fibers\\ Shared telecom infrastructure with WDM traffic\end{tabular}} \\
		& (stable temperature, weak external disturbances)                                                                                       &                                                                                                                                                        \\
		& Dedicated dark fibers with no telecom traffic                                                                                        &                                                                                                                                                        \\ \hline
		\multirow{6}{*}{Key challenges}                                  & \multirow{6}{*}{\begin{tabular}[c]{@{}c@{}}Minimizing device-induced noise\\ Minimizing device-induced non-reciprocity\end{tabular}} & Greater actual link loss                                                                                                                               \\
		&                                                                                                                                      & Temperature fluctuations and vibration                                                                                                                 \\
		&                                                                                                                                      & Calibrating link reciprocity                                                                                                                           \\
		&                                                                                                                                      & Fiber cuts                                                                                                                                             \\
		&                                                                                                                                      & Polarization mode dispersion                                                                                                                           \\
		&                                                                                                                                      & Traffic-induced crosstalk                                                                                                                              \\ \hline
		\multirow{3}{*}{Noise sources}                                   & ASE in amplifiers                                                                                                                    & Environmental phase noise                                                                                                                              \\
		& Electronic device noise                                                                                                              & Power grid instability at remote sites                                                                                                                 \\
		& Residual phase noise from imperfect compensation                                                                                     & Nonlinear effects in high-traffic fibers                                                                                                               \\ \hline
		Implementations                                                  & Explore the limits of system performance                                                                                             & Time-frequency transfer in real-world scenarios                                                                                                        \\ \hline\hline
	\end{tabular}
\end{table*}  \vspace*{-3pt}

As an example, Fig.~\ref{delay-unsuppressed} presents the phase noise spectral density of the free-running 3,000-km link (black curve) alongside the calculated delay-unsuppressed noise (blue curve) derived from Eq.~(\ref{Unsuppressed_Noise_Eq}). The data is from Ref.~\cite{yu_microwave_2024}. It can be seen that in the low-frequency region (below ${1 \mathord{\left/
		{\vphantom {1 {4\tau }}} \right.
		\kern-\nulldelimiterspace} {4\tau }} \approx 15{\rm{Hz}}$, the phase noise cannot be fully suppressed by the compensation system and is fundamentally constrained by the delay-unsuppressed noise. In the high-frequency region (above 15 Hz), phase noise remains entirely uncompensated due to the limited compensation bandwidth imposed by the link delay.

\begin{figure*}[t]
	\centering
	\includegraphics[width= \linewidth]{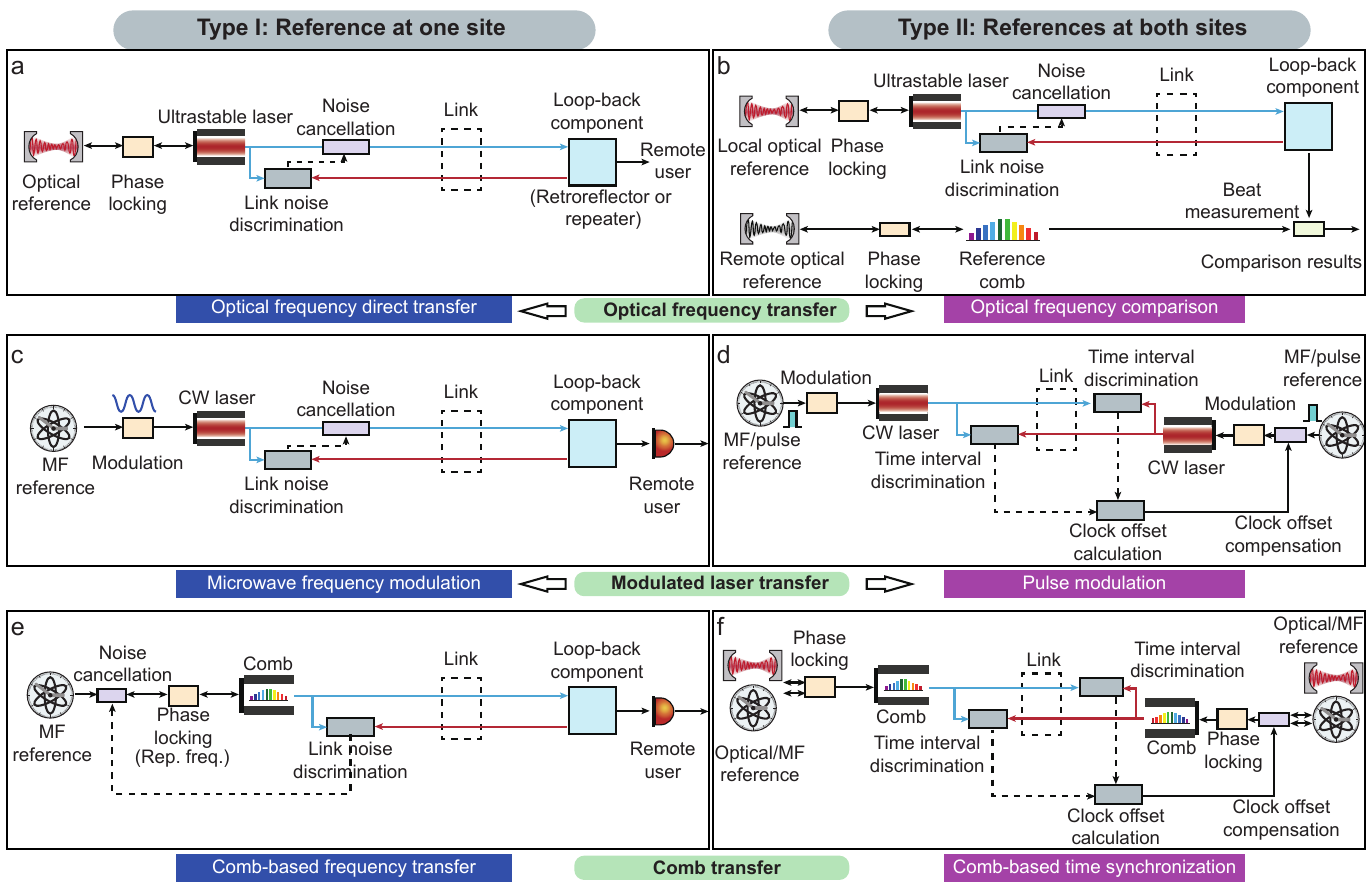}
	\caption{Mainstream transfer methods. The left panel corresponds to the type-I structure where the reference is located at only one site, while the right panel represents the type-II structure where the references are located at both sites. (a) optical frequency transfer; (b) optical frequency comparison; (c) microwave frequency modulation; (d) pulse modulation; (e) comb-based frequency transfer; (f) comb-based time synchronization.
	}\label{Different_methods}
\end{figure*}

\paragraph{\textbf{Practical hurdles in implementations}} Beyond signal quality recovery, temperature fluctuations and phase noise cancellation, real-world deployment faces infrastructural and operational challenges:

\begin{itemize}
	\item \textbf{Power-grid stability.} Remote amplification nodes are relying on stable power supplies. In rural or disaster-prone areas, grid instability can disrupt amplifiers, degrading signal continuity. Solutions like backup batteries or solar power add cost and complexity. The new inline-amplification-free scheme provides a possibility to eliminate the problem of relays~\cite{zhang2024gainbandwidthproductinducedtechnicalboundtime,zhang2024inlineamplificationfreetimetransferutilizing}.
	\item \textbf{Accidental fiber cuts.} Construction, natural disasters, or vandalism frequently sever fiber cables. Redundant paths mitigate this but require costly duplicate infrastructure. Also, after the interruption of time-frequency signals, the recovery of the initial phase is also a critical practical issue that needs to be considered.
	\item \textbf{Telecom traffic interference.} Shared fibers carrying  time-frequency signals risk cross-talk from data traffic. While WDM isolates channels, high traffic loads can still perturb phase stability.
\end{itemize}

Table~\ref{tab:Comparison_lab_field } summarizes the differences between laboratory experiments and field-deployed fiber experiments. These hurdles underscore the gap between laboratory demonstrations and field deployments. Addressing them demands collaboration with telecom providers, robust redundancy designs, and hybrid free-space/fiber fail-safes, which are key steps toward building a resilient global time-frequency network.

\section{MAINSTREAM TRANSFER METHODS}
Figure~\ref{Different_methods} summarizes three mainstream transfer methods: optical frequency transfer, modulated laser transfer, and comb transfer. Based on the accessibility scenarios of optical references, we classify optical time-frequency transfer systems into two categories: Type I (left panel of Fig.~\ref{Different_methods}) and Type II (right panel of Fig.~\ref{Different_methods}).

In Type I configurations, a high-precision reference clock is exclusively available at one site, while the remote site directly employs the transmitted time-frequency signal as its local clock reference. In contrast, Type II systems feature high-precision clock sources at both sites, utilizing time-frequency transfer to achieve inter-site clock comparison or synchronization. Since the presence or absence of a clock at the remote site is primarily determined by its application scenarios, we will proceed to discuss the topic by categorizing according to the three mainstream methods in the following.

\subsection{Optical Frequency Transfer}
\paragraph{\textbf{Method overview}} Optical frequency transfer directly transmits stable optical frequency signals by actively canceling the link noise. The general structure can be seen in Fig.~\ref{Different_methods} a. The optical frequencies are stabilized on optical references such as optical clocks or oscillators. Using an optical frequency reference in lieu of microwaves can significantly enhance the resolution when quantifying phase noise variations of the fiber link. To achieve a low-noise phase reference, the carrier is typically stabilized on a cavity-stabilized ultra-stable laser, whereas high-stability microwave frequency is recovered at the remote site by beating with a stable local optical reference. However, the generated microwave signal by beating with a remote stable optical reference cannot follow the stability of the optical reference in sender.

Alternatively, if high-precision optical references are at both sites, frequency comparision can be acheived through a stablized optical frequency comb, as shown in Fig.~\ref{Different_methods}~b. This structure can also be used to download diverse optical frequencies, microwave frequencies, and timescale signals at the remote site.

\paragraph{\textbf{Historical highlights}}
There is no definitive starting point for research on optical frequency transfers, since the use of optical fibers to transmit laser outputs to adjacent experimental platforms or laboratories has been widespread with the development of optical fiber technology. Before the 1990s, researchers were not particularly concerned about the impact of optical fibers on the quality of laser transmission.

In 1992, Y. Pang \textit{et al.} at the University of Maryland investigated the effects of low-frequency oscillations, vibrations, and acoustic vibrations of optical fibers on optical frequency signals~\cite{Pang:92} and revealed that the quality of laser signals could not be guaranteed after transmission through optical fibers. In 1994, in collaboration with the University of Colorado, NIST proposed the first noise compensation scheme in optical frequency transfer~\cite{Ma:94}. The evaluation scheme for optical frequency transfer was akin to measuring the linewidth of lasers and involves the self-heterodyne technique. The acousto-optic modulator (AOM) was used as the phase compensation component. In 2003, J. Ye \textit{et al.} at the Joint Institute for Laboratory Astrophysics (JILA) used a similar noise compensation scheme, experimented on a 3.45-km optical fiber, and obtained the transfer stability of $3 \times {10^{ - 15}}{\tau ^{-1/2}}$~\cite{Ye:03}. 
In 2007, a novel distance extension method using a phase-coherent transceiver configuration was proposed, effectively extending the transmission distance without excessively sacrificing stability performance~\cite{PhysRevLett.99.153601};
the approach achieved $6 \times {10^{ - 18}}{\tau ^{ - 1/2}}$ for a 7-km link and $2 \times {10^{ - 17}}{\tau ^{ - 1/2}}$ for a 32-km link, which marked a breakthrough with an optimization of three orders of magnitude. Meanwhile, groups at NIST introduced a coherent, frequency-diverse fiber-optic network to provide the potential for the full application of ultra-stable lasers~\cite{coddington_coherent_2007}. Then, an optical amplifier was used to extend the transmission distance to 251 km with the fractional frequency stability of $3 \times {10^{ - 16}}{\tau ^{ - 3/2}}$~\cite{newbury_coherent_2007}.

Notably, long-haul optical frequency transfer has made rapid breakthroughs in the past decade. A 920-km transfer was achieved on the fiber links between Max-Planck-Institut für Quantenoptik (MPQ) and Physikalisch-Technische Bundesanstalt (PTB) using cascaded bidirectional EDFAs, FBAs, and the active noise compensation~\cite{predehl_920-kilometer_2012}. This transfer exhibited a stability of $5 \times {10^{ - 15}}$ for the short term and reached $4 \times {10^{ - 19}}$ for the long term. After the stability analysis method had been improved using frequency counters, more accurate stability results were found over 1,840 km of fiber, following $2.7 \times {10^{ - 15}}{\tau ^{ - 2}}$ and reaching $4 \times {10^{ - 19}}$ on an average time of 100 s. Most recently, ultra-stable lasers were compared through a 2,220-km fiber link using the European metrological optical fiber link network, which connects the National Physical Laboratory (NPL) to PTB via Laboratoire de Physique des Lasers (LPL), LNE-SYRTE, and the University of Strasbourg (UoS). This comparison contributed to various applications of optical clocks such as fundamental research~\cite{schioppo_comparing_2022}.

\subsection{Modulated Laser Transfer}
\paragraph{\textbf{Method overview}} Modulated laser transfer is achieved by modulating a microwave frequency or pulse onto the intensity or phase of continuous-wave (CW) laser for transmission. Modulated frequency signals are typically used with a reference at one site for conducting frequency transfer experiments, whereas modulated pulses usually involve references at both sites to achieve clock synchronization via O-TWFTF, as shown in Fig.~\ref{Different_methods}~c and Fig.~\ref{Different_methods}~d.

In 1980, the Jet Propulsion Laboratory (JPL) of the USA used two connected 1.5-km multi-mode optical fibers to create a 3-km round-trip link~\cite{Lutes1981OpticalFF}. It demonstrated the capability to modulate the output of an amplitude-modulated laser with microwave signals of 20 Hz to 1.2 GHz. This work discussed the transfer using a hydrogen clock and first provided stability of microwave frequency transfer with $2.1 \times {10^{ - 15}}$ at the 100-s averaging time. In addition, three types of phase compensation schemes for microwave frequency transfer were introduced for the first time: feedback compensation, conjugator compensation, and feedforward compensation. Subsequently, most compensation schemes are extensions of these three frameworks. The comparison of difference compensation methods is given in Table~\ref{tab:Comparison_compensation}.

\begin{table*}[t]
	\vspace*{-2pt}
	\centering
	\caption{Comparison of different compensation methods.}
	\label{tab:Comparison_compensation}
	\begin{tabular}{cccc}
		\hline\hline
		Compensation scheme                                                      & Feedforward                                                                                                       & Feedback                                                                      & Conjugator                                                          \\ \hline
		\multirow{2}{*}{Compensation site}                                       & Remote compensation                                                                                               & \multirow{2}{*}{Local compensation}                                           & \multirow{2}{*}{Local compensation}                                 \\
		& \begin{tabular}[c]{@{}c@{}}(Require additional \\ data transmission)\end{tabular}                                 &                                                                               &                                                                     \\ \hline
		\multirow{2}{*}{Compensation stability}                                  & \multirow{2}{*}{\begin{tabular}[c]{@{}c@{}}Dependent on delay and \\ precision of data transmission\end{tabular}} & Best                                                                          & \multirow{2}{*}{High}                                               \\
		&                                                                                                                   & (Especially for short link)                                                   &                                                                     \\ \hline
		\begin{tabular}[c]{@{}c@{}}Possibility of \\ downloadable nodes\end{tabular} & No                                                                                                                & Yes                                                                           & Yes                                                                 \\ \hline
		Complexity and cost                                                      & \begin{tabular}[c]{@{}c@{}}Lowest complexity \\ and cost\end{tabular}                                             & \begin{tabular}[c]{@{}c@{}}Relatively low \\ complexity and cost\end{tabular} & \begin{tabular}[c]{@{}c@{}}High complexity \\ and cost\end{tabular} \\ \hline
		\multirow{2}{*}{Suitable scenarios}                                      & \multirow{2}{*}{Cost limited}                                                                                     & Short link                                                                    & \multirow{2}{*}{Long link}                                          \\ \cline{3-3}
		&                                                                                                                   & Slow temperature drift                                                        &                                                                     \\ \hline\hline
	\end{tabular}
\end{table*}  \vspace*{-3pt}

\textbf{Feedback compensation} is achieved by calculating the phase difference between the round-trip signal and the local reference signal and simultaneously compensating for a phase difference on both forward and backward signals at the local site. 

\textbf{Conjugator compensation} is achieved by shifting the phase of the local loading signal, so that it is always the conjugate of the round-trip signal relative to the zero phase of the reference signal. This step guarantees the phase of the remote signal to remain at zero. 

\textbf{Feedforward compensation} is achieved by calculating the phase difference between the round-trip signal and the reference signal and compensating for phase noise by phase shifting the output signal at the remote site. In practical systems, it is necessary to transmit the phase information after phase discrimination from the transmitter to the receiver and transmit classical digital information using additional channels in fibers, 4G or 5G networks.

\paragraph{\textbf{Historical highlights}} Below we provide a brief overview of the timeline for the three modulation laser compensation methods.

(I) \textbf{Feedback compensation.}
The first type of feedback compensation method is optical feedback compensation by adding a controllable fiber stretcher (or optical delay line) at the local site before transmitting to the fiber link. This approach was first realized in 2000 by the National Astronomical Observatory of Japan~\cite{836302}, which controlled the phase drift of a 100-meter fiber caused by temperature changes through an optical delay control module. Then, JPL implemented a temperature-controlled 40-m optical fiber in a range of 50 $^\circ {\rm{C}}$ to compensate for the delay variations in a buried 16-km optical fiber~\cite{Calhoun2002AS1}. The most significant challenge in the method of temperature feedback fiber length is the gradual change in fiber length with temperature, which makes it difficult to compensate for fast changing noise. To overcome this limitation, the LPL group wound a 15-m optical fiber on a piezoelectric zirconate titanate (PZT) with a 1-km temperature-controlled fiber spool to compensate for both fast and slow phase noise variations in the link. They achieved stability of $1 \times {10^{ - 14}}@1{\rm{s}}$ on an 86-km installed fiber~\cite{PhysRevLett.94.203904,narbonneau_high_2006}, which showed improvements in both distance and stability compared with previous works. In 2010, the LPL group added a dispersion compensating fiber; by increasing the transmitted frequency from 1 to 9.15 GHz, they enhanced the signal-to-noise ratio during phase detection and improved the stability of the 86-km transfer to $1.3 \times {10^{ - 15}}@1{\rm{s}}$~\cite{lopez_high-resolution_2010}, which is the best-reported performance of optical delay line compensation schemes. 

The optical feedback compensation scheme exhibits the best stability at the same transmission distance. Since this scheme directly adjusts the length of the optical fiber, all signals in the same fiber can be simultaneously compensated,
which reduces the need for additional compensation systems, especially for transmitting multiple signals. This method helps reduce the system complexity and costs. However, when the frequency transfer distance increases, the required fiber length and temperature control range must also be increased, which makes it difficult to achieve compensation in long-haul transfer.

The second type of feedback compensation method is electrical feedback compensation. Electrical feedback compensation uses electrical phase shifters at the transmitter to shift both input and round-trip signals by the same phase and ensures a constant phase difference between them. The use of electrical phase shifters can solve the problem of insufficient dynamic range in optical feedback. This approach was first proposed by the AGH University of Science and Technology in 2011~\cite{5658149}. To reduce the impact of asymmetric two-way distributed amplification on long-haul transmission, this group designed a single-path bidirectional EDFA where signals in both directions passed through the same segment of gain fiber. Using this single-path bidirectional EDFA, they achieved a frequency transfer performance of $1.8 \times {10^{ - 13}}@10{\rm{s}}$ over a 420-km link~\cite{liwczyski2013DisseminationOT}. In 2015, the modulated laser transfer over a 2,960-km link was achieved with the stability of $1.9 \times {10^{ - 12}}@1{\rm{s}}$.

(II) \textbf{Conjugator compensation.} 
The emergence of the conjugator compensation scheme predates the use of cables for transmitting microwave-frequency signals. In 1975, JPL first proposed the use of bidirectional cables to transmit microwave-frequency signals and compensate for phase variations during transmission~\cite{macconnell1975microwave}. The conjugator compensation scheme has been a research hotspot due to its large dynamic range and the absence of high-speed acquisition and computation requirements.

In 2007, the National Institute of Information and Communications Technology (NICT) of Japan conducted initial implementations on a 5-km laboratory link~\cite{4319195} and achieved the cascaded transmission of microwave-frequency signals over a 90-km $+$ 114-km link, which reached stability of $2 \times {10^{ - 14}}@1{\rm{s}}$~\cite{5361538}. In 2012, B. Wang \textit{et al.} from Tsinghua University implemented the improved conjugation compensation scheme in an 80-km installed fiber for a 9.1-GHz frequency transfer experiment~\cite{wang_precise_nodate}. This scheme is also known as the “passive compensation" scheme and has been widely used since its proposal. 

The team from Beijing University of Posts and Telecommunications (BUPT) achieved a breakthrough. In 2018, D. Wang \textit{et al.} optimized the noise coefficient of the EDFA, successfully transmitted a 2.4-GHz signal over a 1,007-km laboratory fiber~\cite{wang_stable_2018}, and achieved $1.2 \times {10^{ - 13}}{\tau ^{ - 1/2}}$. In 2021, a dual-phase-locked-loops structure was proposed to optimize the stability of $8.2 \times {10^{ - 14}}@1{\rm{s}}$ and $7.87 \times {10^{ - 17}}@10,000{\rm{s}}$~\cite{liu_ultrastable_2021}. From 2022 to 2023, BUPT and Peking University (PKU) achieved significant milestones in frequency transfer and covered distances of 2,000 and 3,000 km, respectively. BUPT achieved stability of $3.84 \times {10^{ - 14}}@1{\rm{s}}$ and $6.24 \times {10^{ - 15}}@10,000{\rm{s}}$, and PKU achieved $8.8 \times {10^{ - 14}}@1{\rm{s}}$ and $8.4 \times {10^{ - 17}}@10,000{\rm{s}}$~\cite{Gao:23}, which is currently one of the best experimental results in microwave-frequency modulation transfer over long distances.

(III) \textbf{Feedforward compensation.}
To the best of our knowledge, no reported works are using this scheme in modulated laser transfer, although it has been applied in comb transfer~\cite{Chen:15}.

\begin{table*}[]
	\vspace*{-2pt}
	\centering
	\caption{Comparison of mainstream time--frequency transfer technologies. MF: microwave frequency.} 	\label{MainstreamProtocol}
	\begin{threeparttable}
			\begin{tabular}{ccccccc}
			\hline
			\hline
			\textbf{Transfer method}                                                              & \textbf{Carrier}                                                                                   & \textbf{Load$^*$}                                                                          & \textbf{Download}                                                                & \textbf{\begin{tabular}[c]{@{}c@{}}Optical\\ frequency\end{tabular}} & \textbf{\begin{tabular}[c]{@{}c@{}}Microwave\\ frequency\end{tabular}} & \textbf{\begin{tabular}[c]{@{}c@{}}Time\\ scale\end{tabular}} \\ \hline
			\multirow{2}{*}{\begin{tabular}[c]{@{}c@{}}Optical frequency\\ transfer\end{tabular}} & \multirow{2}{*}{\begin{tabular}[c]{@{}c@{}}Ultrastable\\ laser\\ (cavity-stabilized)\end{tabular}} & \multirow{2}{*}{\begin{tabular}[c]{@{}c@{}}Optical freq. \\stabilization\end{tabular}} & \begin{tabular}[c]{@{}c@{}}Optical freq.\\ detection\end{tabular}                & $\checkmark$                                                                  & $\times$                                                                  & $\times$                                                          \\ \cline{4-4}
			&                                                                                                    &                                                                                        & \begin{tabular}[c]{@{}c@{}}Download\\ comb\end{tabular}                          & $\checkmark$                                                                   & $\checkmark$                                                                   & $\checkmark$                                                           \\ \hline
			\multirow{2}{*}{\begin{tabular}[c]{@{}c@{}}Modulated laser\\ transfer\end{tabular}}   & \multirow{2}{*}{CW laser}                                                                          & MF modulation                                                                          & \multirow{2}{*}{\begin{tabular}[c]{@{}c@{}}Intensity/phase\\ demodulation\end{tabular}} &  $\times$                                  & $\checkmark$                      &  $\times$             \\ \cline{3-3}
			&                               & Pulse modulation                                                                       &                                                                                         &  $\times$                                                                   &  $\times$                                                                      & $\checkmark$                                                          \\ \hline
			\multirow{2}{*}{Comb transfer}                                                        & \multirow{2}{*}{\begin{tabular}[c]{@{}c@{}}Optical frequency \\ comb\end{tabular}}                 & \begin{tabular}[c]{@{}c@{}}Mode spacing\\ stabilization\end{tabular}                   & \begin{tabular}[c]{@{}c@{}}Rep. freq.\\ detection\end{tabular}                   & $\times$                                                                 & $\checkmark$                                                                     & $\checkmark$                                                           \\ \cline{3-4}
			&                                                                                                    & \multicolumn{1}{l}{Full stabilization}                                                 & \begin{tabular}[c]{@{}c@{}}Optical freq. \&\\  rep. freq. detection\end{tabular} & $\checkmark$                                                                  & $\checkmark$                                                                     & $\checkmark$                                                           \\ \hline
			\hline
		\end{tabular}
		\begin{tablenotes}
		\item[*]	Stabilization and modulation are achieved by referencing the carrier to the corresponding optical frequency standard (or oscillator) or microwave frequency standard (or oscillator).
		\end{tablenotes}
	\end{threeparttable}

\end{table*} \vspace*{-3pt}

\subsection{Comb Transfer}
\paragraph{\textbf{Method overview}} Comb transfer achieves simultaneous or partial transfer of the optical frequency, microwave frequency, and timescale pulses by directly transmitting the optical frequency comb, which is determined by the characteristics of the comb.

The comb generated by the mode-locked laser contains frequencies of two degrees of freedom: the repetition frequency ${f_{\rm{r}}}$ and the carrier-envelope offset (CEO) frequency ${f_\text{o}}$. Therefore, the frequency of each component is: 
\begin{equation}
	{f_n} = n{f_{\rm{r}}} + {f_\text{o}}.
\end{equation}
As Table~\ref{MainstreamProtocol} shows, using the microwave frequency reference to stabilize mode spacing, the microwave frequency and time-domain pulse envelope (namely microwave conversion timescale) can be transferred. Furthermore, by stabilizing both mode spacing and an optical frequency to the references, multiple optical frequencies, microwave frequencies, and timescale information can be simultaneously transmitted.

Similar to modulated laser transfer schemes, scenarios with a reference clock at one site can be utilized to implement comb-based frequency transfer, while scenarios with reference clocks at both sites enable two-way time comparison and synchronization, as shown in Fig.~\ref{Different_methods}~e and Fig.~\ref{Different_methods}~f. respectively.

\paragraph{\textbf{Historical highlights}} In 2003, T. R. Schibli \textit{et al.} introduced the balanced optical cross-correlation (BOC) technique to directly phase-lock the envelopes between two optical frequency combs~\cite{Schibli:03}. This principle enabled the achievement of a comb synchronization precision of 300 attoseconds. Afterward, long-term femtosecond link stabilization using a single-crystal BOC was demonstrated~\cite{4452578}.

Two technologies are crucial for the comb transfer scheme: linear optical sampling (LOS) and FLOM-PD. In 2003, LOS was proposed~\cite{1253522} and laid the groundwork for sub-picosecond and sub-femtosecond-level time measurements and synchronization~\cite{Coddington:09,coddington_rapid_2009}. In 2012, J. Kim \textit{et al.} from the Korea Advanced Institute of Science and Technology (KAIST) introduced the FLOM-PD scheme. This approach aimed to directly phase-lock an optical frequency comb to a microwave signal and eliminate the phase noise introduced using a photodetector to directly detect the comb. It significantly enhanced the phase-locking precision and synchronization accuracy between the optical frequency comb and the microwave reference signal~\cite{Kim:04,Jung:12}.

In 2004, a fully stabilized optical frequency comb was first used for frequency transfer~\cite{Holman:04}. The signal was transmitted through a buried optical fiber of 6.9 km back to the transmitter with the stability of $2.3 \times {10^{ - 14}}@1{\rm{s}}$. Then, a dispersion compensation fiber was introduced to optimize the comb transfer scheme~\cite{Holman2005RemoteTO}. In 2010, NPL achieved frequency transfer and link noise compensation by fiber stretcher over a 50-km fiber by solely locking the repetition frequency to a hydrogen maser clock in the laboratory~\cite{Marra:10}. This achievement resulted in a stability of $4.6 \times {10^{ - 15}}@1{\rm{s}}$, which was comparable to the performance achieved by modulated laser schemes over similar distances.

In 2015, the research group from PKU used FLOM-PD to lock the repetition frequency of the comb and subsequently used digital feedforward compensation techniques over a 120-km installed fiber to achieve a stability of $8.2 \times {10^{ - 16}}@1{\rm{s}}$~\cite{Chen:15}. In 2024, microwave frequency transfer over a 3,000-km fiber based on optical frequency combs and active noise cancellation was achieved, which demonstrated the effectiveness of using all-optical relays for high-SNR comb transfer.

\begin{figure*}[t]
	\centering
	\includegraphics[width= \linewidth]{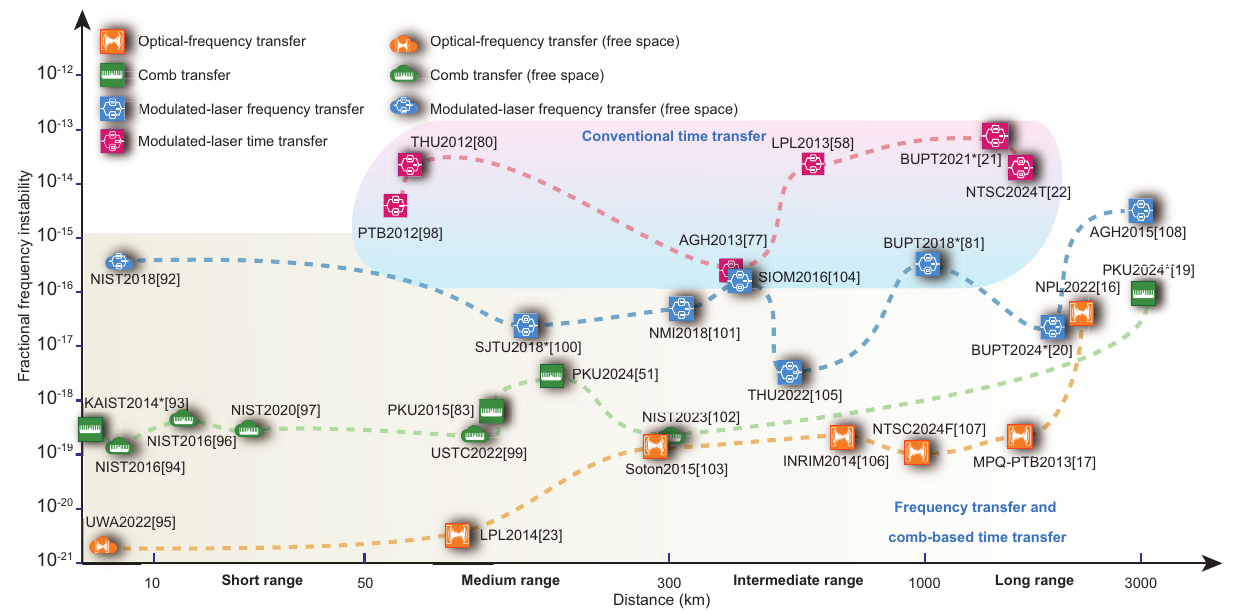}
	\caption{Typical works from different research groups worldwide on O-TWTFT~\cite{Khader:18, Jung:14, deschenes_synchronization_2016, PhysRevLett.128.020801, sinclair_synchronization_2016, bodine_optical_2020, wang_precise_nodate, Rost_2012, Chen:15, shen_free-space_2022, PhysRevA.90.061802, Deng:18, chen_dual-comb-enhanced_2024, He:18, caldwell_quantum-limited_2023, Kim:15, 6702214, liwczyski2013DisseminationOT, Liu:16, 9919221, calonico_high-accuracy_2014, wang_stable_2018, XueDeng:20602, Lin:21, ChinPhysLett.41.064202, schioppo_comparing_2022, Gao:23, droste_optical-frequency_2013, 7138841, yu_microwave_2024}. The squares and clouds denote optical fiber and free space, respectively, according to the transmission medium. Based on the time--frequency signal loading methods, the blue modulator icon represents modulated laser schemes, the green comb icon represents comb transfer schemes, and the orange optical cavity icon represents optical frequency transfer schemes. The star-marked results were obtained from laboratory fiber experiments.}\label{Results}
\end{figure*}

Another benefit of combs is high-precision time synchronization. To overcome the limitation of the picosecond-level uncertainty caused by the photoelectric detection of pulses, in recent years, high-precision time synchronization in free-space and fiber-optic scenarios has been achieved using optical frequency combs by replacing traditional time interval discrimination methods (shown in Fig.~\ref{Different_methods}~f) with newly developed approaches~\cite{chen_dual-comb-enhanced_2024,zhang_hundred-femtosecond-level_2023,yu_time_2023,caldwell_quantum-limited_2023,shen_free-space_2022,bodine_optical_2020,abuduweili_sub-ps_2020,deschenes_synchronization_2016}, such as LOS. 
Femtosecond-level time synchronization can be achieved over hundreds of kilometers in both free-spaces~\cite{caldwell_quantum-limited_2023,shen_free-space_2022} and fiber-optics~\cite{chen_dual-comb-enhanced_2024} scenarios.
This advancement provides a pathway for large-scale networked applications of high-precision clocks.

The characteristics of optical frequency transfer, modulated laser transfer, and comb transfer are also summarized in Table~\ref{MainstreamProtocol}.

\subsection{Comparison of Different Schemes}

Here, we summarize typical works from different research groups worldwide in O-TWTFT~\cite{Khader:18, Jung:14, deschenes_synchronization_2016, PhysRevLett.128.020801, sinclair_synchronization_2016, bodine_optical_2020, wang_precise_nodate, Rost_2012, Chen:15, shen_free-space_2022, PhysRevA.90.061802, Deng:18, chen_dual-comb-enhanced_2024, He:18, caldwell_quantum-limited_2023, Kim:15, 6702214, liwczyski2013DisseminationOT, Liu:16, 9919221, calonico_high-accuracy_2014, wang_stable_2018, XueDeng:20602, Lin:21, ChinPhysLett.41.064202, schioppo_comparing_2022, Gao:23, droste_optical-frequency_2013, 7138841, yu_microwave_2024}, as shown in Fig.~\ref{Results}. The squares and clouds denote optical fiber and free space, respectively, according to the transmission medium. Based on the time--frequency signal loading methods, the blue modulator icon represents modulated laser schemes, the green comb icon represents comb transfer schemes, and the orange optical cavity icon represents optical frequency transfer schemes. All icons are located in the bottom region. The red modulator icon represents works that only focused on the time transfer performance~\cite{Rost_2012,wang_precise_nodate,liwczyski2013DisseminationOT,6702214,Lin:21,ChinPhysLett.41.064202}, which are located in the upper region. Limited by the measurement capability of the electrical pulse, the stability of the traditional modulated laser time transfer is generally lower than frequency transfer. The transmission distance categorization on the horizontal axis is divided into four key regions (especially for frequency transfer):
\begin{itemize}
	\item \textbf{Short range (below tens of kilometers)}~\cite{Khader:18,Jung:14,PhysRevLett.128.020801,deschenes_synchronization_2016,sinclair_synchronization_2016,bodine_optical_2020}. Free-space transfer schemes hold significant advantages, particularly free-space optical frequency transfer schemes~\cite{PhysRevLett.128.020801}, since the impact of free space on the optical carrier is minimal over short distances.
	
	\item \textbf{Medium range (several tens of kilometers to hundreds of kilometers, typically 300 km)}~\cite{shen_free-space_2022,Chen:15,Deng:18,chen_dual-comb-enhanced_2024,caldwell_quantum-limited_2023,Kim:15}. Since atmospheric turbulence limits scramble the optical phase of the received light, two dominant approaches have emerged: using fiber for the optical frequency transfer or using optical pulses to replace the continuously operating optical link in free space.
	
	\item \textbf{Intermediate range (hundreds of kilometers to 1,000 km)}~\cite{He:18,Liu:16,9919221,calonico_high-accuracy_2014,XueDeng:20602,wang_stable_2018}. Line-of-sight constraints in free space and light energy loss restrict the options here. Primarily, cascaded optical fiber transfer schemes are viable.
	
	\item \textbf{Long range (beyond 1,000 km)}~\cite{droste_optical-frequency_2013,Gao:23,schioppo_comparing_2022,yu_microwave_2024,7138841}. The performances of various transfer schemes converge due to temperature fluctuations that govern long-term noise in ultra-long-haul transmission. The unique characteristics of different transfer schemes become indiscernible under these conditions. Therefore, as indicated by Eq.~(\ref{Temp_delay}), temperature control is the key to achieve long-term stability at such long distances.
\end{itemize}

On the vertical axis denoted by the MDEV, we selected the lowest reported value for different integration times ($\tau$) to signify the best performance reported. We overlooked distinctions between short-term and long-term stability to emphasize the technical boundaries achieved by diverse research groups and to streamline comparisons. When works present results in terms of ADEV, we approximated the conversion by multiplying by $1/\sqrt{2}$. For optical comb schemes (green icons), since many of these endeavors simultaneously convey both time and frequency, we selected the frequency performance as metrics in the diagram. For time transfer schemes (red icons), we selected the lowest TDEV values, which are typically found in the literature, and converted them using the formula ${\rm{MDEV}} = \left( {\sqrt 3 {\rm{/}}\tau } \right) \cdot {\rm{TDEV}}$.

It is crucial to note that the star-marked results in Fig.~\ref{Results} were obtained from laboratory fiber experiments. As evident from Table~\ref{tab:Comparison_lab_field }, extending these results to field-deployed fiber scenarios introduces additional technical challenges requiring resolution.

\section{CONCLUSION AND OUTLOOK}\label{sec5}
In conclusion, we have presented the development of the
O-TWTFT over optical fiber, including the characterization of the time--frequency transfer stability, system configuration and key modules, main challenges, and mainstream transfer methods.

The three primary optical time-frequency transfer methods each exhibit distinct characteristics. Optical frequency transfer demonstrates exceptionally high stability, making it ideal for transmitting optical clock signals.
Modulated laser transfer leverages mature optical communication components, offering cost-effectiveness for high-precision microwave clock transfer scenarios. Comb transfer, through its unique transmission carrier, enables simultaneous conveyance of timescale, microwave frequency, and optical frequency signals, providing users with diverse signal formats. These varied methods provide users with flexible options for application. Moreover, rigorous comparison of time-frequency transfer performance requires unified evaluation criteria. Ad hoc metric switching without stated premises invalidates comparative analyses.

The advancing field of fiber-based O-TWTFT necessitates finding a balance between long-distance reach and high precision. The critical goal in establishing extensive time--frequency networks involves selecting suitable transfer methods for distinct application scenarios to optimize both performance and cost-effectiveness.

For prospective applications, the potential merits of optical time-frequency transfer lie in the following aspects:

\paragraph{\textbf{Development of novel time-frequency transfer methods}} New technologies such as repeaterless single-span transfer schemes~\cite{zhang2024gainbandwidthproductinducedtechnicalboundtime,zhang2024inlineamplificationfreetimetransferutilizing} are valuable for future long-haul implementation, since effective relay nodes may not be available in certain special but crucial scenarios such as extremely deficient infrastructures, emergency communications, and those lacking signal regeneration techniques. For large-scale implementations, the shift toward higher integration~\cite{Li:23,10272208,Akatsuka:20} such as chip-based implementations becomes paramount. On-chip integration of critical modules such as light sources and detection is vital for advancing the high-precision time--frequency transfer in the future, particularly for applications in satellites and mobile platforms. Finally, the fundamental assumption of O-TWTFT, which relies on the reciprocity of bidirectional light travel, mandates diligent contemplation of the security of fiber optic time--frequency transmission in intricate networking setups. Minimizing the measurement noise in the link is crucial to enable sensitive detection of variations in fiber delays. This approach can accurately model the time-evolving state of the link and mitigate potential security threats such as delay attacks~\cite{10129234,dai_towards_2020,8357814}.

\paragraph{\textbf{Future applications in fundamental physics and PNT}} The integration of high-precision optical time-frequency transfer over optical fibers promises transformative advancements in both fundamental physics research and PNT systems.

\begin{itemize}
	\item \textbf{Relativity tests}. Networks of synchronized optical clocks could measure gravitational redshift at unprecedented precision~\cite{PhysRevLett.118.221102}. Such experiments might refine general relativity or uncover deviations hinting at new physics.
	
	\item \textbf{Dark matter search}. Future dark matter experiments may deploy distributed detectors requiring high-precision synchronization. Timing coherence deviations in networked atomic clocks serve as probes for topological defect interactions, including domain wall transits~\cite{Nat.Phys.10.933.2014}.
	
	\item \textbf{Environmental sensing}. Coherent fiber networks could monitor earthquake, or environment temperature changes by detecting subtle phase variations of time-frequency signals~\cite{doi:10.1126/science.aat4458}.
	
	\item \textbf{International timekeeping}. Optical clocks, with fractional frequency instabilities approaching $10^{-18}$, could redefine Coordinated Universal Time (UTC). By linking continental-scale optical clocks via ultra-stable fiber networks, UTC could be updated in real time, eliminating averaging delays and reducing uncertainties.
	\item \textbf{Enhancement for satellite navigation}. Fiber-based time-frequency transfer could underpin next-generation navigation by providing ground stations with higher-level timing references, and meanwhile enhancing the resilience of the entire time-frequency network.
\end{itemize}

\paragraph{\textbf{Integrating fiber-based and free-space O-TWTFT: toward global coverage}}
While fiber-based systems excel in stability and precision, free-space optical or satellite links are indispensable for extending time-frequency networks to remote, maritime, or airborne platforms. A hybrid architecture leveraging both technologies could achieve seamless global coverage while balancing performance and accessibility. For instance, in the future, it can be envisioned that a solution combining fiber optics as the transcontinental backbone with satellites as feeder devices will be feasible. Specifically, fiber provides ultra-stable backbone connections between major metrological hubs (e.g., national labs), and satellites or drones act as "feeders" to distribute time-frequency signals to islands, ships, or polar stations lacking fiber infrastructure. This integration is critical for future global clock networks, where fiber ensures bedrock stability for precision applications, while free-space extends reach to underserved regions.

Overall, establishing high-precision, long-distance, highly secure, and chip-scale time--frequency transfer is crucial for advancing metrological capabilities to enable global sharing of high-precision clocks. This progress is expected to seamlessly integrate with existing fiber infrastructures and enable stable and reliable deployments. Such advancements will bring this technology closer to numerous applications in future clock networks and mark a significant step toward improved timekeeping and synchronization across a wide range of applications and industries.

\appendix

\begin{table*}[t]
	\caption{\label{NoiseTable}Typical noise types, together with their resulting Allan variance and modified Allan variance~\cite{Rubiola_2008}.}%
	\begin{tabular*}{\textwidth}{@{\extracolsep\fill}ccc@{\extracolsep\fill}}%
		\toprule
		\textrm{Noise type}&
		\textrm{$\sigma_y^2(\tau)$}&
		\textrm{$\text{mod}~\sigma_y^2(\tau)$}\\
		\colrule
		White PM    & $\frac{3f_\text{H}h_2}{(2\pi)^2}\tau^{-2}$   & $\frac{3f_\text{H}\tau_0h_2}{(2\pi)^2}\tau^{-3}$   \\
		Flicker PM    & $\frac{[1.038+3\ln(2\pi f_\text{H}\tau)]\times h_1}{(2\pi)^2}\tau^{-2}$   & $0.084h_1\tau^{-2}$   \\
		White FM    & $0.5h_0\tau^{-1}$   & $0.25h_0\tau^{-1}$   \\
		Flicker FM     & $2\ln(2)h_{-1}$   & $\frac{27}{20}\ln(2)h_{-1}$   \\
		Random walk FM    & $\frac{(2\pi)^2h_{-2}}{6}\tau$   & $\frac{0.824\times(2\pi)^2h_{-2}}{6}\tau$  \\
		\botrule
	\end{tabular*}
\end{table*}

\section{APPENDIX A: CALCULATIONS OF DIFFERENT METRICS\label{app1}}
Here we present the calculation formula of different metrics metinoned in the section ``characterization of time-frequency transfer stability'' in the main text.
\subsection{Timing Jitter}
\begin{equation}
	{\sigma _{{\rm{jitter}}}} = \sqrt {\int_{{f_{\rm{L}}}}^{{f_{\rm{H}}}} {{S_x}} (f)df} ,
\end{equation}
where $f_\text{L}$ and $f_\text{H}$ represent the low and high frequency cutoffs determined by the filtering properties.

\subsection{Allan Variance, Non-Overlapping}
\begin{equation}
	\scalebox{0.7}{$
		\begin{aligned}
			\sigma_y^2(\tau)&=\frac{1}{2m^2(\lfloor\frac{M}{m}\rfloor-1)}\sum_{j=1}^{\lfloor\frac{M}{m}\rfloor-1}\left[\sum_{i=m(j-1)+1}^{mj}(y_{i+m}-y_i)\right]^2\\
			&=\frac{1}{2m^2\tau_0^2(\lfloor\frac{N-1}{m}\rfloor-1)}\sum_{j=1}^{\lfloor\frac{N-1}{m}\rfloor-1}\left(x_{m(j+1)+1}-2x_{mj+1}+x_{m(j-1)+1}\right)^2,
		\end{aligned}
		$}
	\label{AVAR}
\end{equation}
where $\lfloor\cdot\rfloor$ stands for the floor function, and $m < M/2$, indicating that for data with a total duration of $T$, the maximum theoretical value of $\tau$ is $T/2$.

\subsection{Modified Allan Variance}
\begin{equation}
	\begin{aligned}
		&\text{mod}~\sigma_y^2(\tau)=\frac{1}{2m^4(M-3m+2)}\times\\
		&\sum_{j=1}^{M-3m+2}\left\{\sum_{i=j}^{j+m-1}\left[\sum_{k=i}^{i+m-1}(y_{k+m}-y_k)\right]\right\}^2\\
		&=\frac{1}{2m^4\tau_0^2(N-3m+1)}\times\\
		&\sum_{j=1}^{N-3m+1}\left[\sum_{i=j}^{j+m-1}(x_{i+2m}-2x_{i+m}+x_i)\right]^2.
	\end{aligned}
	\label{MVAR}
\end{equation}

\subsection{Time Variance}
\begin{equation}
	\sigma_x^2(\tau)=\frac{\tau^2}{3}\cdot\text{mod}~\sigma_y^2(\tau).
	\label{TVAR}
\end{equation}

\subsection{Allan Variance, Overlapping}
\begin{equation}
	\label{OAVAR}
	\scalebox{0.87}{$
		\begin{aligned}
			\sigma_y^2(\tau)&=\frac{1}{2m^2(M-2m+1)}\sum_{j=1}^{M-2m+1}\left[\sum_{i=j}^{j+m-1}(y_{i+m}-y_i)\right]^2\\
			&=\frac{1}{2m^2\tau_0^2(N-2m)}\sum_{j=1}^{N-2m}\left(x_{j+2m}-2x_{j+m}+x_j\right)^2.
		\end{aligned}
		$}
\end{equation}

\subsection{Total Variance}
\begin{equation}
	\scalebox{0.88}{$
		\text{TOTVAR}(\tau)=\frac{1}{2m^2\tau_0^2(N-2)}\sum_{i=2}^{N-1}(x_{i-m}^*-2x_i^*+x_{i+m}^*)^2.
		$}
\end{equation}
The newly added data points satisfy $x_{1-j}^* = 2x_1 - x_{1+j}$ and $x_{N+j}^* = 2x_N - x_{N-j}$, where $j = 1, 2, \dots, N-2$. The new sequence $\{x_i^*\}$ extends from $i = 3-N$ to $i = 2N-2$, containing a total of $3N-4$ data points.

\subsection{Theoretical Variance \#1}
\begin{equation}
	\begin{aligned}
		\widehat{\text{Theo1}}(\tau=0.75m\tau_0)=\frac{1}{0.75(N-m)(m\tau_0)^2}\sum_{i=1}^{N-m}\\ \sum_{\delta=0}^{\frac{m}{2}-1}\frac{1}{\frac{m}{2}-\delta}\left[(x_{i+m}-x_{i-\delta+\frac{m}{2}})-(x_{i+\delta+\frac{m}{2}}-x_i)\right]^2,
	\end{aligned}
\end{equation}
where $10 \leq m \leq N-1$ and $m$ is an even number. Furthermore, the bias of $\widehat{\text{Theo1}}$ relative to AVAR is defined with the empirical formula given by
\begin{equation}
	\text{bias}(\tau)\triangleq\frac{\text{AVAR}}{\widehat{\text{Theo1}}}=a+\frac{b}{\tau^c}.
	\label{Theo1bias}
\end{equation}
For different types of noise, the parameters can be found in the reference~\cite{Howe2004}.

\subsection{TheoH}
\begin{equation}
	\text{TheoH}(\tau)\triangleq\left\{
	\begin{aligned}
		&\text{AVAR}(\tau=m\tau_0),~1\le m< \frac{k}{\tau_0} \\
		&\text{TheoBR}(\tau=0.75m\tau_0),\\ 
		&\frac{k}{0.75\tau_0}\le m\le N-1,~m~\text{even}
	\end{aligned}
	\right.
\end{equation}
with
\begin{equation}
	\scalebox{0.95}{$
		\text{TheoBR}\triangleq\left[\frac{1}{n+1}\sum_{i=0}^n\frac{\text{AVAR}(m=9+3i,\tau_0)}{\widehat{\text{Theo1}}(m=12+4i,\tau_0)}\right]\times\widehat{\text{Theo1}}.
		$}
\end{equation}
Here $n=\lfloor\frac{N}{30}-3\rfloor$, and $k$ corresponds to the largest $\tau$ within 10\% of the data run time.

\subsection{Hadamard Variance}
\begin{equation}
	\scalebox{0.8}{$
		\begin{aligned}
			\text{H}~\sigma_y^2 (\tau) &= \frac{1}{6(M-2)}\sum_{i=1}^{M-2} \left( y_{i+2} - 2y_{i+1} + y_i \right)^2 \\
			&=\frac{1}{6m^2\tau_0^2(N-3m)} \sum_{i=1}^{N-3} \left( x_{i+3} - 3x_{i+2} + 3x_{i+1} - x_i \right)^2.
		\end{aligned}
		$}
\end{equation}

\subsection{Structure Function}
The definition of the $N$th-order structure function $D_\psi^{(N)}$ of the random process $\psi(t)$ is
\begin{equation}
	D_\psi^{(N)}=\mathbb{E}\{[\Delta_\tau^N\psi(t)]^2\},
	\label{StructureFunction}
\end{equation}
where the stochastic process $\psi(t)$ may represent either phase or frequency process. Here $\Delta_\tau^N$ stands for the $N$th-order increment, which can be defined based on an iterative approach, namely,
\begin{equation}
	\Delta_\tau^N\psi(t)\triangleq\Delta_\tau^{N-1}[\Delta_\tau\psi(t)].
\end{equation}
Here, $\Delta_\tau$ is the forwards difference operator, and $\tau = m\tau_0$ represents the corresponding integration time. we can obtain through recursion
\begin{equation}
	\Delta^N_\tau x(t)=\sum_{k=0}^N(-1)^kC_N^kx(t+(N-k)\tau),
\end{equation}
where $N>1$.

\section{APPENDIX B: TYPICAL RELATIONSHIP\label{app2}}
Starting from the PSD in the frequency domain, there are methods to derive the aforementioned variances through integration by combining different transfer functions, namely~\cite{Rubiola_2008,handbook2008}
\begin{equation}
	\begin{aligned}
		&\sigma_y^2(\tau)=\int_0^{f_\text{H}}S_y(f)\frac{2\sin^4(\pi f\tau)}{(\pi f\tau)^2}df,\\
		&\text{mod}~\sigma_y^2(\tau)=\int_0^{f_\text{H}}S_y(f)\frac{2\sin^6(\pi f\tau)}{N^4(\pi f\tau_0)^2\sin^2(\pi f\tau_0)}df,\\
		&\sigma_x^2(\tau)=\int_0^{f_\text{H}}S_y(f)\frac{2\tau^2\sin^6(\pi f\tau)}{3N^4(\pi f\tau_0)^2\sin^2(\pi f\tau_0)}df,\\
		&\text{H}~\sigma_y^2(\tau)=2^4\int_0^{f_\text{H}}S_y(f)\frac{\sin^6(\pi\tau f)}{(\pi\tau f)^2}df,
	\end{aligned}
	\label{PSDtoVAR}
\end{equation}
where $f_\text{H}$ denotes the high-bandwidth cutoff. Note that we often use the transfer function ${\left| {H(jf)} \right|^2}$ and combine it with the subscripts to unify the factor of the integral term on the right-hand side of Eq.~(\ref{PSDtoVAR}). Moreover, if we decompose the noise power spectral density into a sum of several specific types of noise,
\begin{equation}
	S_y(f)=\left\{
	\begin{aligned}
		&~~~0,&f\ge f_\text{H} \\
		&\sum_{\alpha=-2}^2h_\alpha f^\alpha,&f<f_\text{H}
	\end{aligned}
	\right.
\end{equation}
then the integration results for these specific types of noise can be directly found in Table~\ref{NoiseTable}~\cite{Rubiola_2008}.

\section{FUNDING}
This work was supported by the National Natural Science Foundation of China (62201012), and the National Hi-Tech Research and Development (863) Program.

\section{AUTHOR CONTRIBUTIONS}
Conceptualization: Z.C., Y.Z., B.L., and H.G.
Data curation: Z.C. and Y.Z.
Formal analysis: Z.C., Y.Z., B.L., and H.G.
Funding acquisition: Z.C. and H.G.
Investigation: Z.C., Y.Z., B.L., and H.G.
Methodology: Z.C., Y.Z., B.L., and H.G.
Project administration: Z.C. and H.G.
Resources: Z.C., B.L., and H.G.
Validation: Z.C., Y.Z., B.L., and H.G.
Visualization: Z.C., Y.Z., B.L., and H.G.
Writing - original draft: Z.C. and Y.Z.
Writing - reviewing and editing: Z.C., Y.Z., B.L., and H.G.
\\
\\

\noindent\textit{\textbf{Conflict of interest statement.}} None declared.

\bibliographystyle{apsrev4-2}
\bibliography{nsr_sample}

%
%
%
%

\end{document}